\documentclass[journal,comst]{IEEEtran}
\usepackage{amsmath}
\usepackage{algorithmic}
\usepackage{array}
\usepackage{amssymb}
\usepackage{acro} 

\DeclareAcronym{IoT}{
    short= IoT, 
    long= Internet-of-Things, 
    tag = abbrev
}

\DeclareAcronym{IoE}{
    short= IoE, 
    long= Internet-of-Everything, 
    tag = abbrev
}

\DeclareAcronym{P2RC}{
    short=P2RC, 
    long= passive-reconfigurable-reflective communication,
    tag = abbrev
}
\DeclareAcronym{ABCm}{
    short=ABCm, 
    long= ambient backscatter communication,
    tag= abbrev
}

\DeclareAcronym{ABCmS}{
    short=ABCmS, 
    long= ambient backscatter communication system,
    tag= abbrev
}

\DeclareAcronym{BBCmS}{
    short=BBCmS, 
    long= bistatic backscatter communication system,
    tag= abbrev
}

\DeclareAcronym{MBCmS}{
    short=MBCmS, 
    long= monostatic backscatter communication system,
    tag= abbrev
}

\DeclareAcronym{RIS}{
    short=RIS, 
    long= reconfigurable intelligent surfaces,
    tag= abbrev
}
\DeclareAcronym{1G}{
  short = 1G,
  long  = first-generation,
  tag = abbrev
}

\DeclareAcronym{5G}{
  short = 5G,
  long  = fifth-generation,
  tag = abbrev
}

\DeclareAcronym{6G}{
  short = 6G,
  long  = sixth-generation,
  tag = abbrev
}
\DeclareAcronym{SRad}{
  short = SRad,
  long  = symbiotic radio,
  tag = abbrev
}

\DeclareAcronym{ISM}{
  short = ISM,
  long  = Industrial Scientific and Medical,
  tag = abbrev
}

\DeclareAcronym{FCC}{
  short = FCC,
  long  = Federal Communication Commission,
  tag = abbrev
}

\DeclareAcronym{ETSI}{
  short = ETSI,
  long  = European Telecommunications Standards Institute,
  tag = abbrev
}

\DeclareAcronym{mmWave}{
  short = mmWave,
  long  = millimeter wave,
  tag = abbrev
}

\DeclareAcronym{THz}{
  short = THz,
  long  = terahertz,
  tag = abbrev
}

\DeclareAcronym{WiFi}{
  short = WiFi,
  long  = wireless fidelity,
  tag = abbrev
}

\DeclareAcronym{TVWS}{
  short = TVWS,
  long  = TV white space,
  tag = abbrev
}

\DeclareAcronym{TV}{
  short = TV,
  long  = television,
  tag = abbrev
}

\DeclareAcronym{FM}{
  short = FM,
  long  = frequency modulation,
  tag = abbrev
}

\DeclareAcronym{LTE}{
  short = LTE,
  long  = long-term-evolution,
  tag = abbrev
}

\DeclareAcronym{SU}{
  short = SU,
  long  = secondary user,
  tag = abbrev
}

\DeclareAcronym{PU}{
  short = PU,
  long  = primary,
  tag = abbrev
}

\DeclareAcronym{CR}{
  short = CR,
  long  = cognitive radio,
  tag = abbev 
}

\DeclareAcronym{BD}{
  short = BD,
  long  = backscatter device,
  tag = abbev 
}

\DeclareAcronym{RFID}{
    short = RFID,
    long  = radio frequency identification,
    tag = abbev
}

\DeclareAcronym{RF}{
  short = RF,
  long  = radio frequency,
  tag = abbev 
}

\DeclareAcronym{eMBB}{
  short = eMBB,
  long  = enhanced mobile broadband,
  tag = abbev 
}

\DeclareAcronym{uRLLC}{
  short = uRLLC,
  long  = ultra-reliable low latency communication,
  tag = abbev 
}

\DeclareAcronym{mMTC}{
  short = mMTC,
  long  = massive machine type communications,
  tag = abbev 
}

\DeclareAcronym{umMTC}{
  short = umMTC,
  long  = ultra massive machine type communications,
  tag = abbev 
}

\DeclareAcronym{Radar}{
  short = Radar,
  long  = radio detection and ranging,
  tag = abbev 
}

\DeclareAcronym{DSA}{
  short = DSA,
  long  = dynamic spectrum access,
  tag = abbev 
}

\DeclareAcronym{SSA}{
  short = SSA,
  long  = static spectrum access,
  tag = abbev 
}

\DeclareAcronym{AASC}{
  short = AASC,
  long  = active-active systems coexistence,
  tag = abbev 
}

\DeclareAcronym{APSC}{
  short = APSC,
  long  = active-passive systems coexistence,
  tag = abbev 
}

\DeclareAcronym{MCS}{
  short = MCS,
  long  = monostatic communication system,
  tag = abbev 
}

\DeclareAcronym{MRS}{
  short = MRS,
  long  = monostatic radar system,
  tag = abbev 
}

\DeclareAcronym{ACS}{
  short = ACS,
  long  = active communication system,
  tag = abbev 
}

\DeclareAcronym{PCS}{
  short = PCS,
  long  = passive communication system,
  tag = abbev 
}

\DeclareAcronym{PRDS}{
  short = PRDS,
  long  = passive radar system,
  tag = abbev 
}

\DeclareAcronym{ARDS}{
  short = ARDS,
  long  = active radar system,
  tag = abbev 
}

\DeclareAcronym{BCmS}{
  short = BCmS,
  long  = backscatter communication system,
  tag = abbev 
}

\DeclareAcronym{ARS}{
  short = ARS,
  long  = active radio system,
  tag = abbev 
}

\DeclareAcronym{PRS}{
  short = PRS,
  long  = passive radio system,
  tag = abbev 
}

\DeclareAcronym{BBC}{
  short = BBC,
  long  = British Broadcasting Corporation,
  tag = abbev 
}

\DeclareAcronym{SCm}{
  short = SCm,
  long  = symbiotic communication,
  tag = abbev 
}

\DeclareAcronym{SRD}{
  short = SRD,
  long  = symbiotic radar,
  tag = abbev 
}

\DeclareAcronym{CSI}{
  short = CSI,
  long  = channel state information,
  tag = abbev 
}

\DeclareAcronym{CDRL}{
  short = CDRL,
  long  = centralized deep reinforcement learning,
  tag = abbev 
}

\DeclareAcronym{DDRL}{
  short = DDRL,
  long  = distributed deep reinforcement learning,
  tag = abbev 
}

\DeclareAcronym{NOMA}{
  short = NOMA,
  long  = non-orthogobal multiple access,
  tag = abbev 
}

\DeclareAcronym{FD}{
  short = FD,
  long  = full-duplex,
  tag = abbev 
}

\DeclareAcronym{MIMO}{
  short = MIMO,
  long  = multiple-input multiple-output,
  tag = abbev 
}

\DeclareAcronym{LIS}{
  short = LIS,
  long  = large-intelligent surface,
  tag = abbev 
}

\DeclareAcronym{UAV}{
  short = UAV,
  long  = unmanned aerial vehilce,
  tag = abbev 
}

\DeclareAcronym{GPS}{
  short = GPS,
  long  = global positioning system,
  tag = abbev 
}

\DeclareAcronym{BR}{
  short = BR,
  long  = backscatter receiver,
  tag = abbev 
}

\DeclareAcronym{AT}{
  short = AT,
  long  = active transmitter,
  tag = abbev 
}

\DeclareAcronym{AR}{
  short = AR,
  long  = active receiver,
  tag = abbev 
}

\DeclareAcronym{OOK}{
  short = OOK,
  long  = on-off keying,
  tag = abbev 
}

\DeclareAcronym{SINR}{
  short = SINR,
  long  = signal-to-interference-plus-noise-ratio,
  tag = abbev 
}

\DeclareAcronym{LAA-LTE}{
  short = LAA-LTE,
  long  = lincensed assisted access long term evolution,
  tag = abbev 
}

\DeclareAcronym{SSaC}{
  short = SSaC,
  long  = symbiotic sensing and communication,
  tag = abbev 
}

\usepackage{hyperref} 
\usepackage{graphicx}
\usepackage{multirow}
\usepackage[table,xcdraw]{xcolor}
\usepackage[utf8]{inputenc}
\usepackage{pifont}
\newcommand{\cmark}{\ding{51}}%
\newcommand{\xmark}{\ding{55}}
\usepackage{cite}
\usepackage{graphicx}
\begin{document}

\title{Survey on Symbiotic Radio: A Paradigm Shift in Spectrum Sharing and Coexistence \thanks{This work has been submitted to the IEEE Journal for possible publication. Copyright may be transferred without notice, after which this version may no longer be accessible.}}

\author{Muhammad Bilal Janjua,
        H\"{u}seyin Arslan,~\IEEEmembership{Fellow,~IEEE}
\thanks{M. B. Janjua is with the Department of Electrical and
Electronics Engineering, Istanbul Medipol University, 34810 Istanbul, Turkey e-mail: (muhammad.janjua@std.medipol.edu.tr).}
\thanks{H. Arslan is with the Department of Electrical and Electronics
Engineering, Istanbul Medipol University, 34810 Istanbul, Turkey, and also
with the Department of Electrical Engineering, University of South Florida,
Tampa, FL 33620 USA (e-mail: huseyinarslan@medipol.edu.tr).}
}

%
\maketitle
\begin{abstract}
\Ac{6G} of mobile communication aims to connect this world digitally through green communication networks that provide secure, ubiquitous, and unlimited connectivity in an attempt to improve the overall quality of life. The driving force behind the development of these networks is the rapid evolution of \ac{IoT}, which has stimulated the proliferation of wireless applications in health, education, agriculture, utilities, etc. However, these applications are accompanied by the deployment of a massive number of \ac{IoT} devices that require a significant radio spectrum for wireless connectivity. \ac{IoT} devices usually have low data rate requirements and limited power provision but desirably a long life. Recently, the development of passive radio systems has opened new paradigms of spectrum sharing and coexistence. These systems utilize the radio resources and infrastructure of the active radio systems to perform their functionalities. By enabling the dependent coexistence, a new technology named \ac{SRad} enables the symbiotic relationships between the different radio systems ranging from mutual benefits or competition in terms of sharing the resources, in particular for \ac{IoT} devices. This survey first provides the motivation for dependent coexistence and background of spectrum sharing through coexistence along with existing literature. Then, it describes the active and passive radio systems definition and a brief overview. Afterward, the history of symbiosis and the role of \ac{SRad} technology in spectrum sharing and coexistence are defined while focusing on symbiotic communication. Lastly, we discuss research challenges, future directions, and applications scenarios.
 
\end{abstract}


\section{Introduction}

    \Acl{6G}  has gained attention in the research and engineering community to connect this world digitally and develop future societies, where everything is working smartly such as living, healthcare, education, etc., \cite{dang2020should}. The driving force behind this aim is \ac{IoT}, which first connected the machines, then humans, and now everything to the Internet \ac{IoE}. If the wireless devices continue to grow at this rapid pace, the number will reach approximately 12.3 billion by 2023 as predicted by Cisco  \cite{cisco2018cisco}. 

    Previous generations of wireless communication have offered several services such as voice, text message, multimedia, \ac{eMBB}, \ac{uRLLC}, \ac{mMTC}, during the development period from \ac{1G} to \ac{5G} to facilitate the user with new experiences \cite{yazar20206g}. However, in \ac{6G} it is planned to go beyond the communication services and incorporate the \ac{Radar} functionalities to the communication systems to achieve accurate wireless sensing and localization \cite{Wang2021sym}. These services will support the development of diverse future applications such as autonomous vehicles, holographic teleportation, extended reality, industrial automation, etc. Even though a lot of research efforts are going on for the realization of these technologies, yet a major bottleneck is the limited radio spectrum, which is insufficient to support the surge of \ac{IoT} devices and sensors coming along \cite{akyildiz20206g}. A many-fold increase in the \ac{RF} spectrum is needed to provide connectivity to such a massive number of devices.

    Radio spectrum scarcity or the lack of spectrum issue has been coming along since the time of Guglielmo Marconi when his transatlantic transmission occupied the entire radio spectrum. As the competitors tried to use the same radio frequencies, their transmission encountered severe interference \cite{cooper2010scarcity}. In earlier times,  the radio spectrum was utilized in broadcasting, wireless telegraphy, and \ac{Radar}. The evolution of technologies and standardization under strong regulations have made it possible to use it for different applications related to military, emergency services, and cellular systems, etc. However, continuously growing applications of wireless systems with changing users' interest from \ac{FM} radio, \ac{TV} to social media after the convergence of Internet and wireless communication technologies have changed the course of spectrum utilization. This sudden change has intensified the spectrum scarcity problem and made the electromagnetic spectrum a precious natural resource like other natural resources such as water, minerals, etc., \cite{ryan2005treating}. Therefore, rules and regulations have been implemented on spectrum utilization to maximize the benefits and reduce the adverse effects, e.g., radio spectrum pollution \cite{Russell2018}. 
    
    At the beginning of radio technology development, exclusive licensing was used to allocate the spectrum considering the signal interference issues. The proliferation of wireless devices and accelerating uses makes it clear that the exclusive licensing strategy is an inefficient way of spectrum utilization \cite{Bhattarai2016}. To ensure the efficient utilization of spectrum and avoiding the interference between radio systems; two spectrum accessing policies are defined that are \ac{SSA} and \ac{DSA}. In \ac{SSA}, the access is either licensed, rule-based or unlicensed. For licensed-based access, users buy the exclusive property rights of the radio spectrum from the regulatory authorities for specific applications or services. In the rule-based access, certain conditions must be satisfied for accessing the radio spectrum, such as fee payment, transmit power levels, and unoccupied channels. Lastly, the license-exempt or unlicensed access is initiated by the United States \ac{FCC} in the \ac{ISM} bands, which allows every technology to access the spectrum with equal rights under some basic regulations \cite{Voicu2019}. On the other hand, \ac{DSA} comes under the concept of \ac{CR}, which is defined because not all the radio spectrum is occupied all the time, and white-spaces/spectrum holes can be utilized by the \acp{SU} \cite{Maloku2018}. These regulations and allocations vary between countries according to their legislative bodies, geo-locations, operating conditions, and technological developments; however, the common objective is to achieve the optimum utilization and to overcome the shortfall of spectrum resources \cite{Papadias2020}.
            
\begin{table*}[h!]
\centering
\caption{LIST of Abbreviations}
\begin{tabular}{|l|l|l|l|}
\hline
1G                    & First-Generation                                & LTE                   & Long-Term-Evolution                           \\ \hline
5G                    & Fifth Generation                                & MBCmS                 & Monostatic Backscatter Communication   System \\ \hline
6G                    & Sixth Generation                                & MIMO                  & Multiple-Input Multiple-Output                \\ \hline
ABCm                  & Ambient Backscatter Communication               & mMTC                  & Massive Machine Type Communications           \\ \hline
ABCmS                 & Ambient Backscatter Communication   System      & mmWave                & Millimeter Wave                               \\ \hline
ACS                   & Active Communication System                     & NOMA                  & Non-Orthogonal Multiple Access                \\ \hline
APSC                  & Active-Passive Systems Coexistence              & OOK                   & On-off   Keying                               \\ \hline
AR                    & Active   Receiver                               & PCS                   & Passive Communication System                  \\ \hline
ARDS                  & Active Radar System                             & PRDS                  & Passive Radar System                          \\ \hline
ARS                   & Active Radio System                             & PRS                   & Passive Radio System                          \\ \hline
AT                    & Active Transmitter                              & PU                    & Primary                                       \\ \hline
BBC                   & British Broadcasting Corporation                & Radar                 & Radio Detection and Ranging                   \\ \hline
BBCmS                 & Bistatic Backscatter Communication   System     & RF                    & Radio Frequency                               \\ \hline
BCmS                  & Backscatter Communication System                & RFID                  & Radio Frequency Identification                \\ \hline
BD                    & Backscatter Device                              & RIS                   & Reconfigurable Intelligent Surfaces           \\ \hline
CDRL                  & Centralized Deep Reinforcement Learning         & SCm                   & Symbiotic Communication                       \\ \hline
CR                    & Cognitive Radio                                 & SINR                  & Signal-to-Interference-Plus-Noise-Ratio       \\ \hline
CSI                   & Channel State Information                       & SRad                  & Symbiotic Radio                               \\ \hline
DDRL                  & Distributed Deep Reinforcement Learning         & SRD                   & Symbiotic Radar                               \\ \hline
DSA                   & Dynamic Spectrum Access                         & SSA                   & Static Spectrum Access                        \\ \hline
eMBB                  & Enhanced Mobile Broadband                       & SSaC                  & Symbiotic   Sensing and Communication         \\ \hline
FCC                   & Federal Communication Commission                & SU                    & Secondary User                                \\ \hline
FD                    & Full-Duplex                                     & THz                   & Terahertz                                     \\ \hline
GPS                   & Global Positioning System                       & TV                    & Television                                    \\ \hline
IoE                   & Internet-Of-Everything                          & TVWS                  & TV   White Space                              \\ \hline
IoT                   & Internet-Of-Things                              & UAV                   & Unmanned Aerial Vehicle                       \\ \hline
ISM                   & Industrial Scientific and Medical               & uRLLC                 & Ultra-Reliable Low Latency   Communication    \\ \hline
LAA-LTE               & Lincensed Assisted   Access Long Term Evolution & WiFi                  & Wireless Fidelity                             \\ \hline
LIS                   & Large-Intelligent Surface                       &                       &                                               \\ \hline
\end{tabular}
\label{table:1}
\end{table*} 
    \begin{table*}[ht]
\caption{Comparison with Surveys in Literature}
\begin{center}
\begin{tabular}{l|cc|cc|cc|cc}
\hline 
\hline
 & \multicolumn{2}{c}{\textbf{Active Systems}} \vline& \multicolumn{2}{c}{\textbf{Passive Systems}} \vline& \multicolumn{2}{c}{\textbf{Spectrum Regulations}} \vline& \multicolumn{2}{c}{\textbf{coexistence}} \\ \cline{2-9}
\multirow{-2}{*}{\textbf{Reference}} 
 & \textbf{Communication}  & \textbf{Radar}  & \textbf{Communication} & \textbf{Radar}  & \textbf{Licensed}         
 & \textbf{Unlicensed}     & \textbf{Active-Passive}   \\ \hline \hline
 This survey
 & \cmark & \cmark    & \cmark  & \cmark   & \cmark & \cmark  & \cmark \\\hline 
\cite{Feng2020,Zheng2019}                                                                   & \cmark & \cmark & {\color{red}\xmark}   & \cmark & \cmark & \cmark  & \cmark \\ \hline 
\cite{Mazahir2021,Bhattarai2016,Choi2020,Voicu2019,Labib2017,Han2016}                                              & \cmark & \cmark    & {\color{red}\xmark}   & {\color{red}\xmark}   & \cmark & \cmark & {\color{red}\xmark} \\ \hline 
\cite{Naik2018,Labib2016}                                                                        & \cmark & \cmark & {\color{red}\xmark}   & {\color{red}\xmark}   & {\color{red}\xmark} & \cmark & {\color{red}\xmark} \\ \hline 

 \cite{Tarek2020,MamadouMamadou2020,Mousa2020,Slamnik-Krijestorac2020,Alejandrino2019,Gupta2019,Masmoudi2019,Hu2018,Amjad2018,Zhang2017a,Zhang2017,Amjad2017,Xia2018,Oyewobi2017,Tehrani2016} 
 & \cmark & {\color{red}\xmark}    & {\color{red}\xmark}   & {\color{red}\xmark}   & \cmark & \cmark & {\color{red}\xmark} \\ \hline 
\cite{Attiah2020,Peng2019,Hoyhtya2017}

& \cmark & {\color{red}\xmark}    & {\color{red}\xmark}   & {\color{red}\xmark}   & \cmark & {\color{red}\xmark} & {\color{red}\xmark} \\ \hline 
\cite{Manekiya2020,Mustafa2019,Zinno2018,Wang2017Ce,BenHafaiedh2018,Bajracharya2018}                                                  & \cmark & {\color{red}\xmark}    & {\color{red}\xmark}   & {\color{red}\xmark}   & {\color{red}\xmark} & \cmark & {\color{red}\xmark}\\ \hline 
\end{tabular}
\end{center}
\label{table:2}
\end{table*}

            Different approaches have been considered to accommodate upcoming surge of \ac{IoT} devices and high spectrum demands, including allocation of new bands (i.e., \ac{mmWave} and \ac{THz}) \cite{Han2019}, re-allocation of legacy licensed spectrum bands \cite{Rosston2014}, \ac{CR} \cite{Tarek2020}, and coexistence of wireless technologies \cite{Voicu2019}. The first two approaches require either innovations and technological enhancements, or transfer of property rights and clearing the existing users, which are exorbitant and time-consuming, respectively  \cite{Rosston2014}. However, \ac{CR} is a more convenient approach than earlier ones in terms of cost and complexity as well as in enhancing the spectrum efficiency of under-utilized bands, where the number of users using the spectrum is less than the available resources. In this case, \ac{CR} users share the under-utilized part of the spectrum as \acp{SU} and enhance the spectrum utilization. However, \ac{CR} systems can create unavoidable interference to the primary systems if not operated under defined regulations. Due to the interference concerns of \acp{CR} in spectrum sharing, their adoption in real-time applications is limited so far \cite{Liang2021}. 
            The coexistence of technologies overcomes the spectrum scarcity issue, where different wireless systems can collaborate to share and access the resources. When two or more systems can share the same band independently while not affecting the performance of the other co-existing system. Several coexistence approaches have been proposed in the literature  such as \ac{WiFi} and \ac{TV} in \ac{TVWS} \cite{Maloku2018}, integrated non-terrestrial and satellite systems \cite{Peng2019}, \ac{WiFi} and \ac{LTE} in unlicensed 5-GHz band \cite{Wang2017Ce}, narrow band \ac{IoT} and \ac{LTE} \cite{Zhang2019}, and communication and radar systems \cite{Labib2017}. However, both the \ac{CR} and aforementioned wireless co-existing systems are \ac{ARS}, which require an \ac{RF} source for signal generation at the transmitter. Although these systems improve the spectral efficiency; however, the active radio systems are not suitable for low power \ac{IoT} devices. Active signal generation and transmission require a significant amount of power that reduces the battery life of \ac{IoT} devices. 
            
 \subsection{Motivation and Background}       

            Unlike the \ac{ARS}, there is another class of radio systems i.e., \acp{PRS}, which do not require a dedicated \ac{RF} source and utilize the radio waves generated by the \ac{ARS} (e.g., FM/\ac{TV}, \ac{WiFi}, cellular systems) termed as illuminators of opportunity (also known as ambient signals). The passive communication through ambient signals is commonly known as \ac{ABCm} in the literature \cite{Liu2013, VanHuynh2018}. Unlike the \ac{ARS} spectrum sharing mechanisms, \ac{PRS} can share both the unlicensed and licensed spectrum without creating significant interference \cite{Guo2019a}. Furthermore, \ac{PRS} not only shares spectrum but also infrastructure of \ac{ARS} i.e., \ac{RF} source. Although \ac{PRS} are spectrum and energy-efficient compared to \ac{ARS}, their complexity is high and performance is low \cite{Griffiths2017,Long2020}. One of the main reasons for performance limitations is that \ac{PRS} can not control or modify the ambient signals according to their requirements. However, the limitations of both \ac{ARS} and \ac{PRS} can be overcome through interaction, and inter-specific associations within these systems for radio resource sharing \cite{Liang2021, Wang2021sym}.     

          In a biological ecosystem, dissimilar organisms share an environment and interact in different ways.  Anton de Bary noticed this behavior and called it `symbiosis' means `living together' \cite{de1879}. He defines symbioses as ``associations between two different species of organisms for food, shelter, or protection." The different ways of interspecific association for resource sharing between organisms are defined by the symbiotic relationships, and each organism is called a symbiont. Furthermore, symbiosis can be obligatory or facultative based on the dependence of the organisms. If two symbionts are so close to each other that one or both cannot live without each other then the symbiosis is obligatory. Otherwise, if both can survive independently but can also be in symbiosis by choice then this type of symbiosis is facultative \cite{Liang2021}.
           
          A similar analogy can be applied to radio ecosystem for resource sharing between two dissimilar systems, i.e., \ac{ARS} and \ac{PRS}, where symbiosis represents the coexistence, symbiotic relationships define the ways of radio resource sharing between systems and the relevant radio system is called \acl{SRad} (\ac{SRad}) \cite{Liang2021}. Moreover, coexistence between radio systems can be dependent (obligatory) or independent (facultative) according to the dependence of systems. If one radio system cannot survive without the other, then this type of coexistence may be termed as dependent coexistence, e.g., the coexistence of \ac{ARS} and \ac{PRS}. On the other hand, in independent coexistence, the radio systems can perform their functions independently but choose to coexist, e.g., \ac{WiFi} and \ac{LAA-LTE} \cite{tan2019qos}. Besides, \ac{SRad} systems can achieve collective objectives as well as individual objectives through mutualistic or competition-based symbiotic relationships. Thus, \ac{SRad} can enhance the performance of both \ac{ARS} and \ac{PRS} through dependent and independent coexistence with efficient radio resource sharing \cite{Guo2019a,Long2020,Han2021}. 

        \subsection{Related Prior Literature}
        Some several tutorials and surveys exist in the literature on spectrum sharing and coexistence. A classification of previous surveys based on system types i.e., active or passive, spectrum band regulations i.e., licensed or unlicensed, and co-existing systems is provided in Table \ref{table:1}. For the explanatory purpose, active and passive systems are sub-categorized into communication and radar systems. While considering the coexistence scenarios, only \ac{APSC} is considered, where one system is active and the other is passive, irrespective of either it is for communication or radar. 
        In \cite{Feng2020, Zheng2019}, authors review the coexistence of active communication and passive radar systems in licensed and unlicensed bands. Overview of active communication and radar systems in both licensed and unlicensed bands is given in  \cite{Mazahir2021, Bhattarai2016, Choi2020, Voicu2019, Labib2017, Han2016} and only in the unlicensed band is discussed in \cite{Naik2018, Labib2016}. Considering the active communication systems coexistence in both licensed and unlicensed, only licensed and only unlicensed bands is discussed in \cite{Tarek2020,MamadouMamadou2020,Mousa2020,Slamnik-Krijestorac2020,Alejandrino2019,Gupta2019,Masmoudi2019,Hu2018,Amjad2018,Zhang2017a,Zhang2017,Amjad2017,Xia2018,Oyewobi2017,Tehrani2016};  \cite{Attiah2020,Peng2019,Hoyhtya2017}; and \cite{Manekiya2020,Mustafa2019,Zinno2018,Wang2017Ce,BenHafaiedh2018,Bajracharya2018}, respectively.
        Although surveys considering all the scenarios are presented in Table \ref{table:1} for completeness, the studies relevant to this work are given in \cite{ Feng2020, Zheng2019}. 
        The authors in \cite{Feng2020} provide a survey on joint communication and radar systems from wireless communication and radar sensing perspectives. Further, they discuss three techniques namely coexistence, codesign, and cooperation for the development of a joint radar communication system considering both active as well as passive radars. The overview of waveform design and signal processing techniques for joint radar and communication is given in \cite{Zheng2019}. The differences between active and passive radar systems and their coexistence mechanism are also discussed along with their applications and use case scenarios.

        \subsection{Contributions}
        The contributions of this survey to the literature are listed below:
    
        \begin{itemize}
            \item To the best of the authors' knowledge, this is the first survey on \ac{SRad} systems that study the spectrum sharing concept in wireless radio systems, leveraging the symbiosis concept. In line with this, the history of symbiosis is described along with applications in wireless systems.
          
            \item The dependence of \acp{PRS} on \acp{ARS} is highlighted and the different types of symbiotic relationships are explored ranging from mutual benefit for both to competition for the sharing of the radio resources, in particular the \ac{RF} spectrum.
        
            \item We investigate the \ac{SCm} and \ac{SRad} paradigms for \ac{IoT} devices and networks through symbiosis between \ac{PRS} and \ac{ARS}.
            \item Lastly, we explore the applications of \ac{SRad} systems for \ac{6G} from human and machine-centric perspectives and highlight open research problems to direct future research in this area.
        \end{itemize}

 \begin{figure}[ht]
\centering
\includegraphics[width=\linewidth]{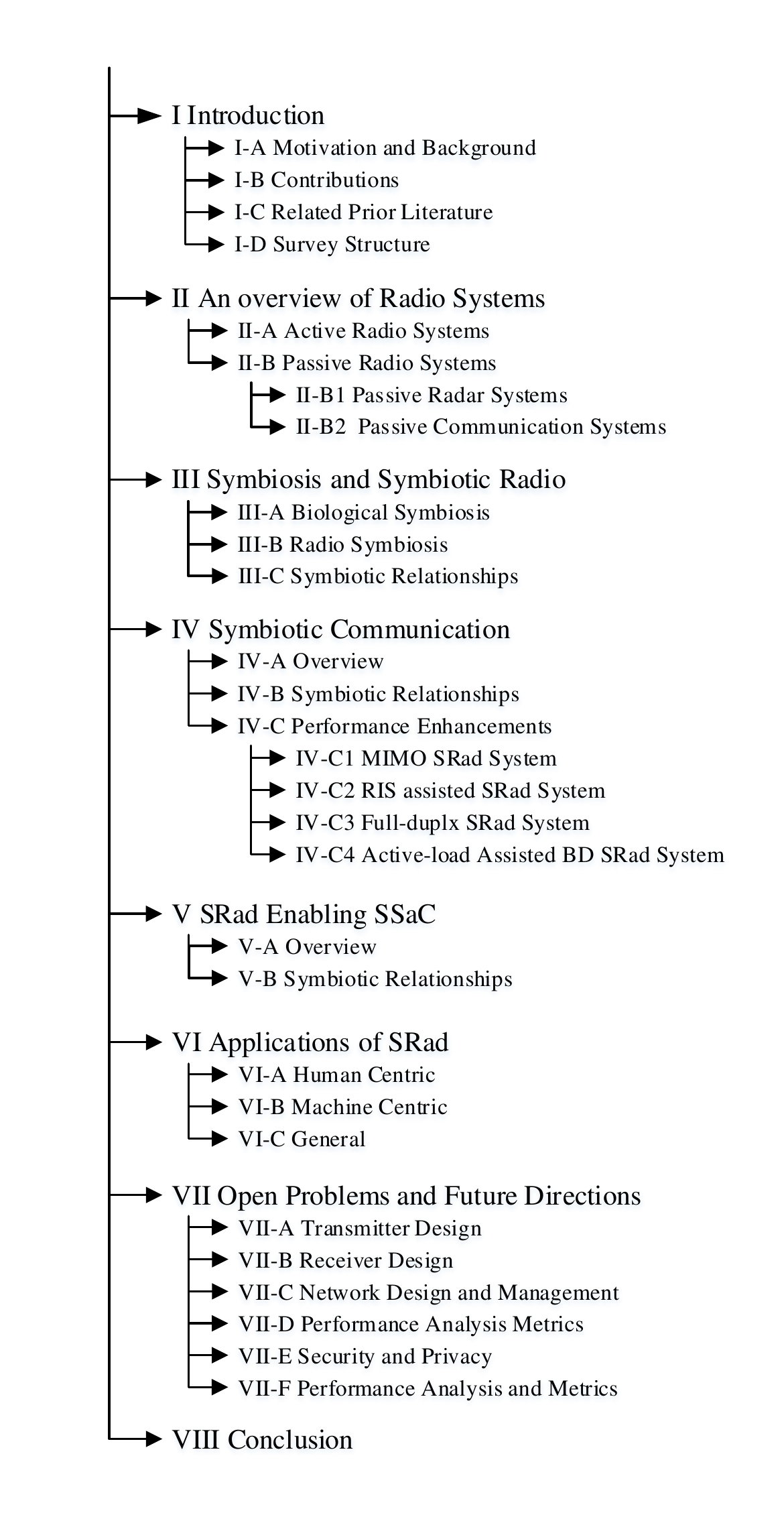}
\caption{The organization of survey.}
\label{fig:Fig1}
\end{figure} 
        \subsection{Survey Structure}
        The organization of the survey is outlined in Fig. \ref{fig:Fig1}. Section II provides an overview of the active and passive radio systems. Details on symbiosis are provided in section III, and different symbiotic radios for spectrum sharing with detailed discussions on symbiotic communication and symbiotic radar are given in section IV. Section V presents the potential applications of \ac{SRad} systems. The critical challenges and future directions for further research and development of \ac{SRad} are provided in section VI. Lastly, the conclusion is provided in section VII. A complete list of abbreviations is given in Table \ref{table:2}.
        
\section{An Overview of Radio Systems}
\begin{figure*}[ht]
\centering
\includegraphics[width=\textwidth]{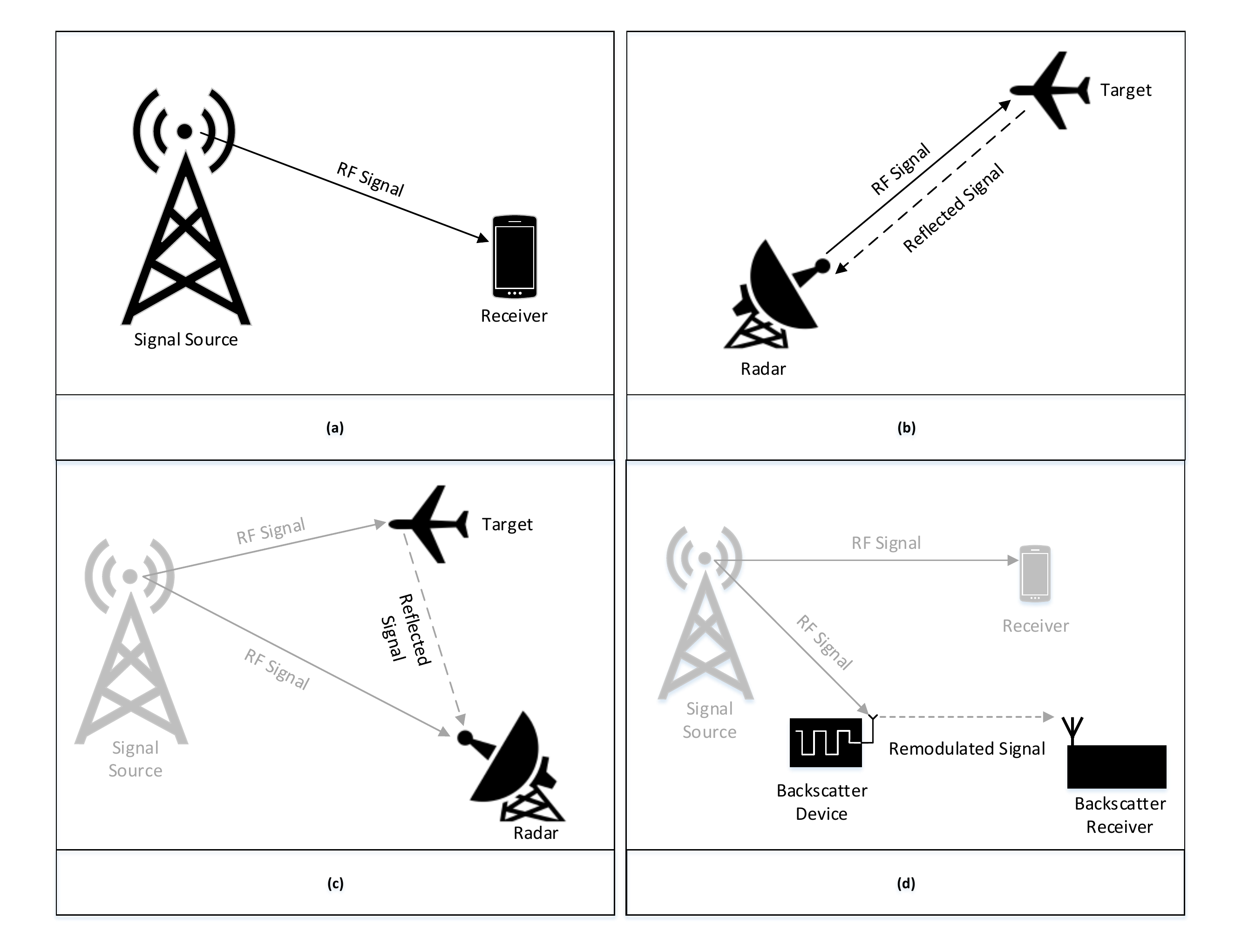}
\caption{Illustration of different radio systems: (a) \ac{ACS}, (b) \ac{ARDS}, (c) \ac{PRDS}, (d) \ac{PCS}.}
\label{fig:Fig2}
\end{figure*}

    This section gives a general overview of active and passive radio systems. Starting from the definition and brief overview of \ac{ARS}, then discuss the \ac{PRS} and its different types. Additionally, communication and radar systems are discussed as use case scenarios for both systems, which are shown in Fig. \ref{fig:Fig2}.  

    \subsection{Active Radio Systems}

        A radio system that generates \ac{RF} signals at the transmitter for information transfer and extracts the information from the transmitted signal at the receiver. A general system for radio signal transmission consists of two parts; transmitter A and receiver B. A sends the information by transmitting a radio signal in the air, where B receives the signal and extracts the information. The radio signal can be an information modulated signal or a continuous wave signal depending on the application. For instance, an \ac{ACS} modulates the carrier i.e., \ac{RF} signal, to send the information to the receiver as shown in Fig. \ref{fig:Fig2} (a). On the other hand, \ac{ARDS} shown in Fig. \ref{fig:Fig2} (b), which contains an \ac{ARDS} transceiver that radiates the radio wave in the air to detect the presence of the target and its related information. \ac{ARDS} can be considered as a special case of \ac{ACS}, where the target sends its information unknowingly by reflecting the radio wave radiated by the \ac{ARDS} transceiver. Furthermore, active signal generation and transmission are necessary for \ac{ARS} either it is a communication system or a \ac{Radar} system. The architecture of the \ac{Radar} system is different from the communication systems because an \ac{ARDS} is a monostatic system, in which the transmitter and the receiver are collocated and a single antenna is used for transmission of \ac{RF} signals and reception of reflected signals. The common examples of \ac{ARS} in communication are FM, \ac{TV}, cellular, \ac{WiFi}, \ac{UAV} etc., and in \ac{Radar} systems are mapping, earth monitoring, navigational, etc.
   
        \subsection{Passive Radio Systems}

        A radio system that reflects/backscatters \ac{RF} signals at transmitter for information transfer and extracts the information from the reflected/backscattered signals at the receiver. The use of \ac{PRS} based on backscattering principle can be dated back to the development of passive radar technology used in World War II. When passive radio reflector was mounted on the allied aircraft to backscatter the signal transmitted by the home radar with better illumination than enemy aircraft \cite{chawla2007}. Afterward, the idea of reflected power communication was proposed by Harry Stockman in 1948, who contributed to the development of \ac{RFID} technologies \cite{stockman1948}. However, a completely passive \ac{RFID} system that is operated and controlled by the illuminated signals, was demonstrated by Alfred Koelle in 1975 \cite{koelle1975}. Since then, the \ac{RFID} technology has been deployed in various fields such as medical, business, logistics, and so on \cite{landt2005}.
        
        \ac{PRS} can be a \ac{PRDS} used in military applications or a \ac{PCS} to provide the connectivity to a simple \ac{IoT} device or sensor. A \ac{PRDS} is shown in Fig. \ref{fig:Fig2}(c), where \ac{Radar} receiver utilizes the radio waves radiated by ambient \ac{ARS} for target detection instead of generating a dedicated signal \cite{Griffiths2017}. On the other hand, in \ac{PCS}, the ambient signals of \ac{ARS} are remodulated and backscattered by a \ac{BD} i.e., tag to send the information to the receiver as shown in Fig. \ref{fig:Fig2}(d). Furthermore, \acp{PRS} can utilize the ambient signals of \ac{RF} sources belong to \ac{ARS}, from the Bluetooth device through to space-based satellites as well as everything and anything in between. However, in practice, only a few ambient sources are favorable for the \ac{PRS} because they are not designed to support these systems \cite{Kuschel2019passive}.
\begin{figure*}
    \centering
    \includegraphics[scale=0.60]{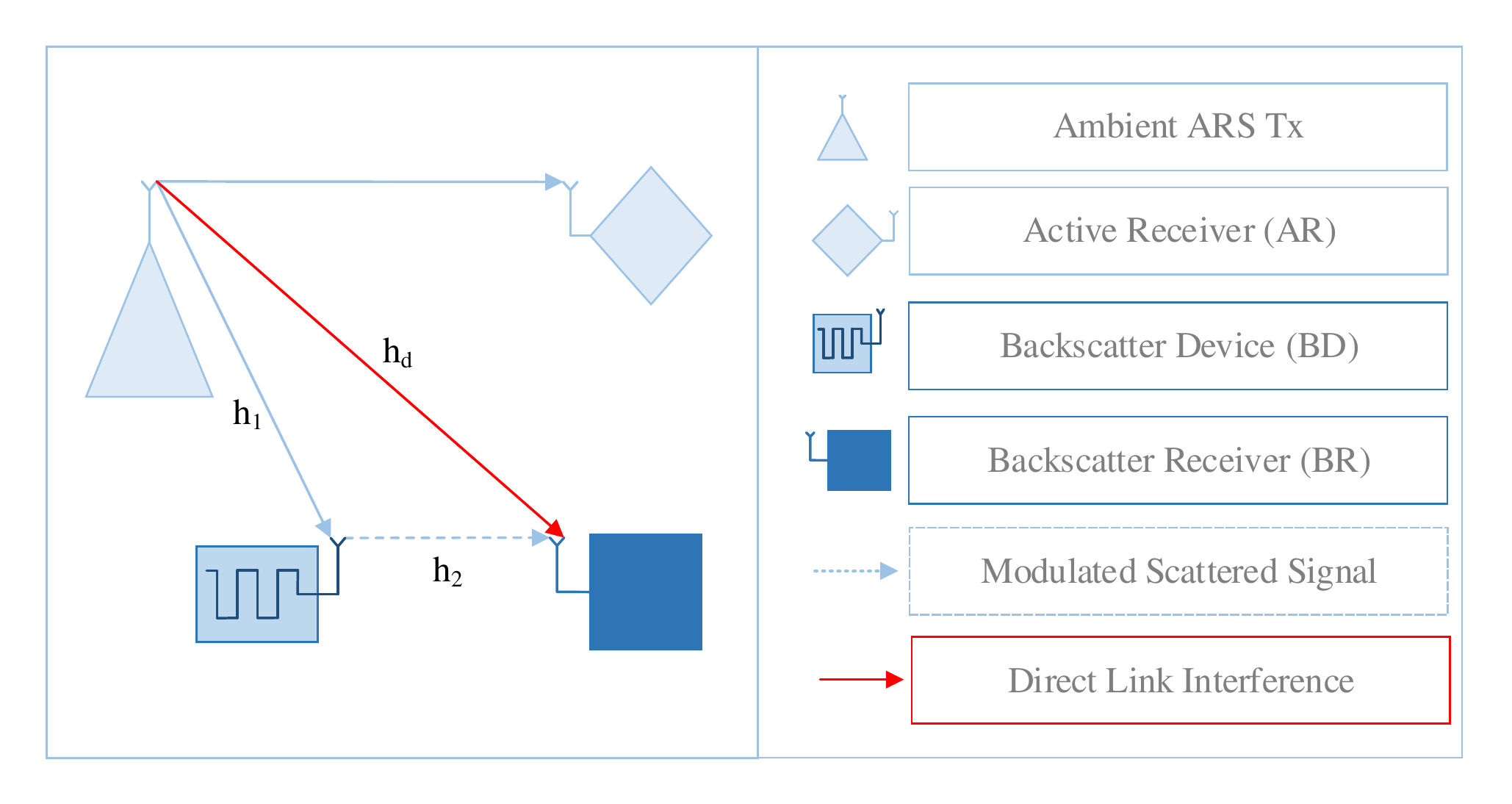}
    \caption{Illustration of conventional \acs{ABCmS} with single \ac{BD}.}
    \label{fig:fig3}
\end{figure*}

        \subsubsection{Passive Radar Systems}
        The first \ac{PRS} was a \ac{PRDS} designed in 1935 by Robert Watson-Watt, which was able to detect the aircraft at a distance of 12 km using the radio signals of \ac{BBC} shortwave transmitter \cite{jones1988technical, Kuschel2010Passiveradar}. After world war II, long-lasting research on \ac{PRDS} started for its advantages that include covertness, ability to operate in the lower frequency bands, the capability of anti-stealth detection, spectral efficiency, and so on \cite{Kuschel2019passive}. \ac{PRDS} are bistatic radars, and they detect their target of interest exploiting ambient \ac{RF} signals reflected by the target. Thus, additional spectrum resources are not needed for \acp{PRDS}, which makes them spectral efficient compared to \acp{ARDS}. However, the ambient radio signals are generated by \acp{ARS} and unknown to the passive radar receiver. The ambient signal has to be measured initially to cross-correlate with the target reflected signal to detect and track the target. The bistatic range and Doppler shift are measured from the time difference of arrival and frequency difference between the direct signal (also called reference signal) coming from \ac{ARS} and the signal reflected off the target. A \ac{PRDS} can also have multiple ambient \ac{ARS} sources or multiple receivers for reception, these types of radars are called multi-static radars. Although \ac{RF} sources of several ambient \ac{ARS} are available, whose signal can be used in \acp{PRDS}, a good \ac{RF} source is the one that provides the continuous signal for a longer time duration \cite{Kuschel2019passive}.  
\newline
        \subsubsection{Passive Communication Systems}
        Similar to \ac{PRDS}, the signals of opportunity (also known as ambient signals) have been used to enable passive communication for low-power sensors and \ac{IoT} devices. These systems are known as \acl{ABCmS}s (\acp{ABCmS}), where \ac{BD} modulate its data over the ambient signals by switching its antenna \cite{Jin-PingNiu2019}. Although \ac{PCS} term has not been used particularly for \ac{ABCmS} but we prefer to use it due to its generality and similarity to \ac{PRDS}. The terms \ac{PCS} and \ac{ABCmS} are used interchangeably throughout the paper. The main idea of \ac{ABCmS} comes from the backscatter communication systems, where a \ac{BD} modulate and scatter the incident signal to transmit its data \cite{chawla2007}. Backscatter communication systems are designed with three configurations of \ac{RF} signal source and receiver: a) \ac{MBCmS}; b) \ac{BBCmS}; c) \ac{ABCmS} \cite{VanHuynh2018}.

                A \ac{PRS} system in which communication is done by reflecting the continuous wave signal transmitted by a dedicated \ac{RF} source, which is located with the receiver in the same device. An \ac{RFID} is the most common example of a \ac{MBCmS}, which consists of a reader and an \ac{RFID} tag i.e., \ac{BD}. Both the \ac{RF} source and the \ac{BR} are located in the reader when the reader wants to communicate to the tag, \ac{RF} source transmits the \ac{RF} signals to activate the tag. Then, the tag backscatters the signal to the reader. Where the receiver gets the information from this signal. Besides, tags are battery-less devices, which harvest the energy from the received signals before transmission. Besides, semi-passive tags with batteries are also used in \ac{RFID} to perform internal functions but not \ac{RF} signal generation or transmission.
                
                In \ac{MBCmS}, as the \ac{RF} source and receiver are collocated, the signal suffers from round trip path loss. Consequently, the performance of the system degrades due to the doubly near-far problem especially when the distance between the tag and the reader is large \cite{Jin-PingNiu2019}. Several techniques have been proposed to enhance the performance of \ac{MBCmS} such as multi-antenna \cite{Griffin2008}, channel modelling and estimation \cite{Kim2003}, anti-collision schemes \cite{Joshi2008}, coding schemes \cite{Simon2006} etc. Although \ac{MBCmS} have a wide range of applications such as positioning, patient identification, tracking, etc., \cite{Munoz-Ausecha2021}. Yet, its applications are limited in long-range and high data rate applications.

                A \ac{PRS} in which communication is done by reflecting the continuous wave signals transmitted by dedicated \ac{RF} source located separately from the receiver is termed as \ac{BBCmS}. A simple \ac{BBCmS} consists of an \ac{RF} source, a \ac{BD}, and a \ac{BR}. The \ac{RF} source generates a signal to enable the communication in \ac{BBCmS}, which is located at a distance from the \ac{BD} but separated from the receiver. The isolation of \ac{RF} source helps in reducing the round-trip path loss as in \ac{MBCmS}, leading to the long-range communication \cite{Kimionis2014}. However, the increase in the range comes at the additional cost for the deployment of \ac{RF} sources. \ac{BBCmS} have applications in \ac{IoT}, wireless sensor networks, surveillance and security \cite{Khan2020, Kampianakis2014, Xu2018}.    
 
                 In an \ac{ABCmS} system communication is done by remodulating and backscattering the modulated ambient \ac{RF} signals, transmitted by \acp{ARS} such as \ac{WiFi} access points, cellular base stations, \ac{TV} towers, etc. Different terms have been used in the literature for \ac{ABCmS} such as FM backscatter, WiFi backscatter, BackFi, HitchHike, etc., \cite{Wang2017FM, bharadia2015, Zhang2016HH}. Nonetheless, all the aforementioned backscatter systems share the common principle of using the ambient radio signals for communication as shown in Fig. \ref{fig:fig3}. In \ac{ABCmS}  system cost is reduced, which is required for the deployment of dedicated \ac{RF} sources in \ac{BBCmS}. Allocation of the frequency spectrum is also not required in \ac{ABCmS} as it shares the same spectrum with the \acp{ARS}  \cite{Liu2013}. To transmit the data, \ac{BD} first harvests the energy from the \ac{RF} signal then remodulates it as a carrier by switching its antenna to reflecting and non-reflecting states to transmit 1 and 0, respectively, or otherwise. On the other hand, the \ac{BR} receives two signals: one is reflected signal from \ac{BD} and the second is direct link signal from the \ac{RF} source. The direct link signal is strong compared to \ac{BD} signal and creates the interference. Besides, an ambient radio signal is unknown to \ac{BR}, which makes the \ac{BD} signal detection a challenging problem for \ac{ABCmS}. Different methods have been proposed in the literature to efficiently detect the \ac{BD} signal in the presence of interference from direct link \ac{RF} signal \cite{Tao2019, Tao2018, Qian2017, Qian2017a, Kang2017}. Further, several other works have been proposed to improve \ac{ABCmS} performance through system design \cite{Liu2013}, coding \cite{Liu2017coding}, and application of multiple antennas \cite{Tao2019antenna,Zhao2019}. 
                
                \textbf{Fundamentals of \ac{PCS}}: A simple \ac{PCS} such as \ac{ABCmS} consists of a transmitter i.e., \ac{BD}, and a receiver i.e., \ac{BR}. Different from other backscattering communication systems, \ac{ABCmS} does not have a dedicated \ac{RF} source and utilize ambient signal transmitted by an \ac{ARS}. The \ac{ARS} can be any ambient system with its own transmitter and receivers e.g., an indoor \ac{WiFi} system to provide the internet connectivity to cell phone users. For ease of understanding, we denote the \ac{RF} source and receiver of \ac{ARS} as \ac{AT} and \ac{AR}, respectively. As in Fig. \ref{fig:fig3}, the channel gains of the links between \ac{AT} and \ac{BD}, \ac{AT} and \ac{BR}, and \ac{BD} and \ac{BR} are represented as $h_d$, $h_1$, and $h_2$, respectively. The AT transmits an information modulated signal $x$, and the signal received at the \ac{BD} represented as $y_{bd}$ is given by
                \begin{equation}
                    y_{bd}= \sqrt{p}h_1x+n,
                \end{equation}
                where $p$ is the transmit power and $n$ is the noise, which is negligible if the \ac{BD} only uses the a passive components \cite{fuschini2008analytical}. However, an active load can also be used at the \ac{BD} for the amplification of incident signal, in that case the contribution of noise is not negligible \cite{Long2020a}. When \ac{BD} receives the signal transmitted by \ac{AT}, it switches the antenna between different reflecting states to re-modulate the signal with its information bits $b$. Each reflecting state defined by a reflection coefficient $\Gamma_j$ is expressed as 
                \begin{equation}\label{eq1}
                    \alpha_j =  \frac{Z_j-Z^{*}_{sc}}{Z_{j}+Z_{a}},
                \end{equation}
                where $z_j$ is the load impedance, $Z^{*}_{sc}$ is the antenna impedance due to structural component, $j$ represents the number of reflecting states, and $*$ is complex conjugate. Thus, time varying $\alpha_j$ can be obtained by changing the load impedance as given by $\phi(t) = \alpha_i$. Furthermore, the $\alpha_i$ can be used to design the constellation set of a modulation scheme for \ac{BD}. For instance, in case of \ac{OOK} two states are required to represent '0' and '1' i.e., reflecting and non-reflecting states. The signal at the output of the \ac{BD} $z_{bd}$ is given by \cite{Jin-PingNiu2019}
                \begin{equation}\label{eq2}
                    z_{bd} = \alpha_j y_{bd},
                \end{equation}
                where $\alpha$ is the reflection efficiency of the \ac{BD} antenna and its value if between 0 and 1.  
                
                Suppose the \ac{BR} samples the signal at Nyquist-information rate of the \ac{AT} signal, which are given by
                \begin{equation}\label{eq3}
                    y_{br}[m] = x_a[m]+x_b[m]+w[m], 
                \end{equation}
                where $y_{br}[m]$ are the samples of the received signal,  $x_a[m]=\sqrt{p}h_dx[m]$ is the signal coming directly from \ac{AT}, $x_{bd}[m]=\sqrt{p}\alpha_j h_1 h_2 z_{bd}[m]$ is the modulated \ac{BD} signal, and $w_{BR}[n]$ represents the noise. The average power of the $M$ received samples can be computed at the \ac{BR} as follows\cite{Liu2013}:
                \begin{equation}\label{eq4}
                    \frac{1}{M} \sum_{i=1}^{M} {|y_{br}[m]|}^2 = \frac{1}{M} \sum_{i=1}^{M}{|x_a[m]+x_b[M]+w[m]|}^2 
                \end{equation}
                The signal $x[m]$ is uncorrelated with noise $w[m]$ and $z_{bd}$ can take only two values j=0 and j=1 in the case of \ac{OOK}. Then, (\ref{eq4}) can be rewritten as follows:
                \begin{equation} \label{eq5}
                \small
                    \frac{1}{M} \sum_{i=1}^{M} {|y_{br}[m]|}^2 = \frac{{|1+\alpha z_{bd}|}^2}{M} \sum_{i=1}^{M}{|x[m]|}^2+\frac{1}{M}\sum_{i=1}^{M}{w[m]|}^2
                \end{equation}
                where the average power of received \ac{AT} signal is $P_{at}=\frac{1}{M} \sum_{i=1}^{M}{|x[m]|}^2$, when \ac{BD} is in reflecting state i.e., $z_{bd}=0$ and non-reflecting i.e., $z_{bd}=1$ states, then the average power at \ac{BR} is $P_{at}$ and ${|1+\alpha z_{bd}|}^2 P_{at}$, respectively. Now, \ac{BR} can decode the \ac{BD} data based on the average power of the signal using a common digital receiver. However, an analog-to-digital converter is required at \ac{BR} to sample the received signal, which consumes a significant amount of power and not suitable for low power systems \cite{VanHuynh2018}. 
\begin{figure*}[ht]
    \centering
    \includegraphics[width=\textwidth] {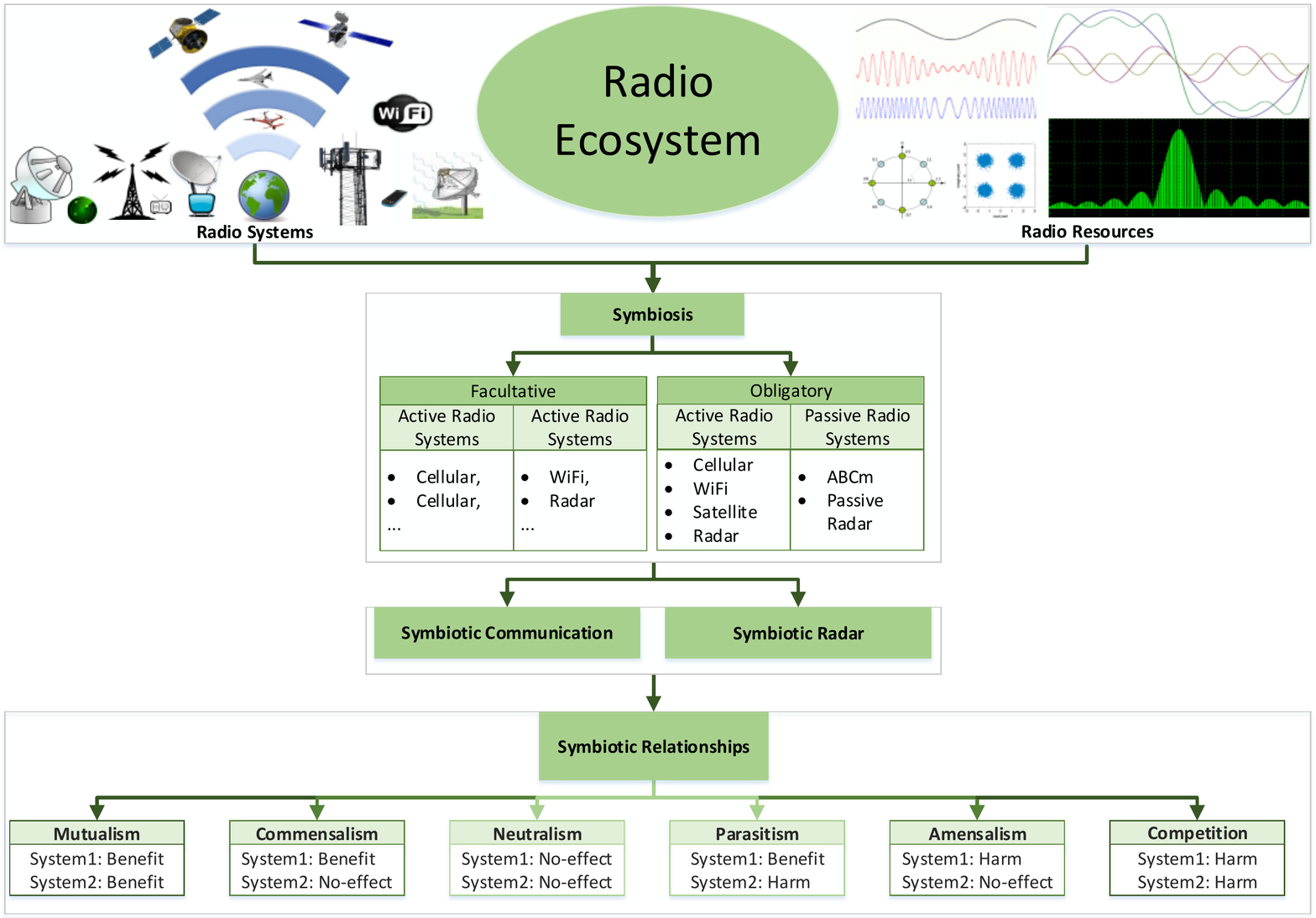}
    \caption{Symbiosis in the radio ecosystem.}
    \label{fig:Fig4}
\end{figure*}     

                \textbf{Discussions:}
                In the conventional design, \ac{ABCmS} share the spectrum with \ac{ARS} without any cooperation which results in major problems, few of them are outlined below: 
                \begin{itemize}
                    \item As \ac{ABCmS} share the spectrum with \ac{ARS}, the signal that comes directly from \ac{AT} is a strong signal compared to \ac{BD} signal and treated as interference at  \ac{BR} signal. Consequently, the transmission rate performance of \ac{ABCm} degrades severely due the direct link interference \cite{wang2016ambient, Guo2019d}. 
                    \item Due to lack of cooperation between the \ac{ABCmS} and \ac{ARS} channel estimation information is unknown at the \ac{BR}, alongside the synchronization is also not possible. Therefore, blind signal detection algorithms are adopted to decode the signal at \ac{BR} but these algorithms are not suitable for low power \ac{BR} due to their high complexity issues \cite{Qian2017a}.
                    \item The ambient signals transmitted by \ac{ARS} are designed independently from the \ac{ABCmS} and have signal characteristics not suitable for passive \ac{BD}. For instance, passive \ac{BD} requires a power-optimized waveform that transmits a maximum amount of power during energy harvesting \cite{trotter2009power}.  
                \end{itemize}

\section{Symbiosis and Symbiotic Radio}
        In this section, we give a brief overview of the radio ecosystem and its relevance to biological symbiosis. Then, we discuss the symbiotic radio along with its types. Lastly, symbiotic relationships are discussed for resource sharing between different radio systems in symbiosis. A general diagram of symbiosis in the radio ecosystem is shown in Fig. \ref{fig:Fig4}.
        
        \subsection{Biological Symbiosis}
        A natural ecosystem consists of two types of components, namely biotic and abiotic. Biotic components include all the living organisms such as animals, plants, bacteria, etc., and abiotic components refer to non-living physical resources that affect the ecosystem such as water, soil, minerals, etc. In a biological ecosystem, multiple biotic components live together and have different associations while sharing physical resources that can be for survival or facultative. In 1879, a pathologist named Anton de Bary coined the term `symbiosis' which means `living together for interspecific associations or interactions among dissimilar organisms \cite{Martin2012}. He defines symbiosis as ``interspecific associations,  or  symbioses,  occur  when  two  different  species of organisms depend on each other for food, shelter, or protection.” The organism that takes part in the symbiosis is called symbiont and the ways in which two or more symbionts interact for resources sharing are defined by symbiotic relationships. The symbiosis can be facultative and obligatory as per the dependence of the symbionts on each other. If one or two of the symbionts completely depend on each other and cannot survive without symbiosis, this type of symbiosis is called obligatory symbiosis. On the other hand, if both symbionts can survive independently but also engage in symbiosis, then this type of symbiosis is termed facultative or optional symbiosis. In each symbiosis, symbionts can have different symbiotic relationships either to achieve the common objectives or individual objectives. The possible symbiotic relationships include but are not limited to mutualism, commensalism, parasitism, neutralism, amensalim, and competition \cite{montesinos2003plant}. 
        
        \subsection{Radio Symbiosis}
        The radio ecosystem is somehow not different than the biological ecosystem. It has biotic components, i.e., radio systems, and abiotic components, i.e., radio resources. Radio systems are categorized into \ac{ARS} and \ac{PRS}, which utilize the radio resources such as frequency, space, time, etc., to perform their functionalities e.g., communication, \ac{Radar}. The current radio ecosystem is evolving rapidly with the proliferation of wireless applications; however, the radio resources particularly the spectrum is limited and insufficient to accommodate the needs of radio systems. Even the \ac{ARS} with exclusive spectrum access, e.g., the ones with licensed spectrum, are unable to meet their stringent requirements. Therefore, they are trying to access the other radio spectrum resources, available to everyone on an equal basis, i.e., unlicensed spectrum. This solution is not feasible because the unlicensed spectrum is becoming populated with wireless devices in each passing and going to face drastic interference issues in the near future. 
        
        In order to meet the radio resource requirements, different radio systems can engage in symbiosis for intelligent radio resource sharing. Symbiosis is analogous to the coexistence between two dissimilar systems, where each radio system called \ac{SRad}, have inter-specific associations and interaction with other \acp{SRad} for sharing the resources. According to the dependencies of \acp{SRad}, the coexistence can be categorized as dependent or independent, similar to obligatory or facultative symbiosis, respectively. In dependent coexistence, one or both \acp{SRad} depend on each other for sharing the radio resources e.g., coexistence between \ac{PRS} and \ac{ARS}, where \ac{PRS} depends on \ac{ARS} for radio resources and cannot survive alone \cite{Kuschel2019passive, VanHuynh2018}. However, the example of independent coexistence is the coexistence between two \acp{ARS}, where both can also survive separately. 
        As the radio systems provide a wide range of functionalities, the \acp{SRad} systems can be designed from radio systems that provide the same radio service e.g., communication only, or radio systems that provide different services e.g, \ac{Radar} and communication. An example of former \acp{SRad} is \ac{SCm}, while \ac{SRD} is an example of the latter. In \ac{SCm}, different radio systems share the radio resources to provide communication services \cite{Liang2021}. However, in \ac{SRD} or \ac{SSaC} communication radio systems and \ac{Radar} and communication radio systems share the radio resources  \cite{Wang2021sym,Mishra2014,Mishra2015}. In this survey, we only focus on \ac{SCm} systems with dependent coexistence and provide brief discussions on \ac{SRD} systems for the sake of completeness.
        
     \subsection{\ac{SRad} Enabling Symbiotic Relationships}
        Similar to biological symbiosis, \acp{SRad} can have different symbiotic relationships irrespective of the type of symbiosis to meet the common or individual objectives from radio resource sharing. Definitions of possible symbiotic relationships are given below:

        \begin{itemize}
            \item \textit{Mutualism:}
             In a mutualistic relationship, radio systems in symbiosis share the radio resources and benefit each other to improve their performances and evolve together. 
             \item \textit{Commensalism:}
            When two radio systems are in a commensal relationship, one system gets benefits by sharing the radio resources, but the other gets no benefit or harm.
            \item \textit{Parasitism:}
            In a parasitic relationship, one system gets benefits by sharing radio resources but it harms the other system in the association.
            \item \textit{Neutralism:}
            This symbiotic relationship defines the association where both systems share the resources independently but no one gets benefits or harm from the association.
            \item \textit{Ammensalism:}
            In an amensal relationship one system when sharing the radio resources with other systems, harms the performance of the other system; however, it gets no benefits or harm from this association.
            \item \textit{Competition:}
            This type of symbiotic relationship defines the association in which both systems compete with each other in sharing the radio resources and affects each other's performance negatively.
        \end{itemize}

\begin{figure*}[ht]
    \centering
    \includegraphics[width=\textwidth] {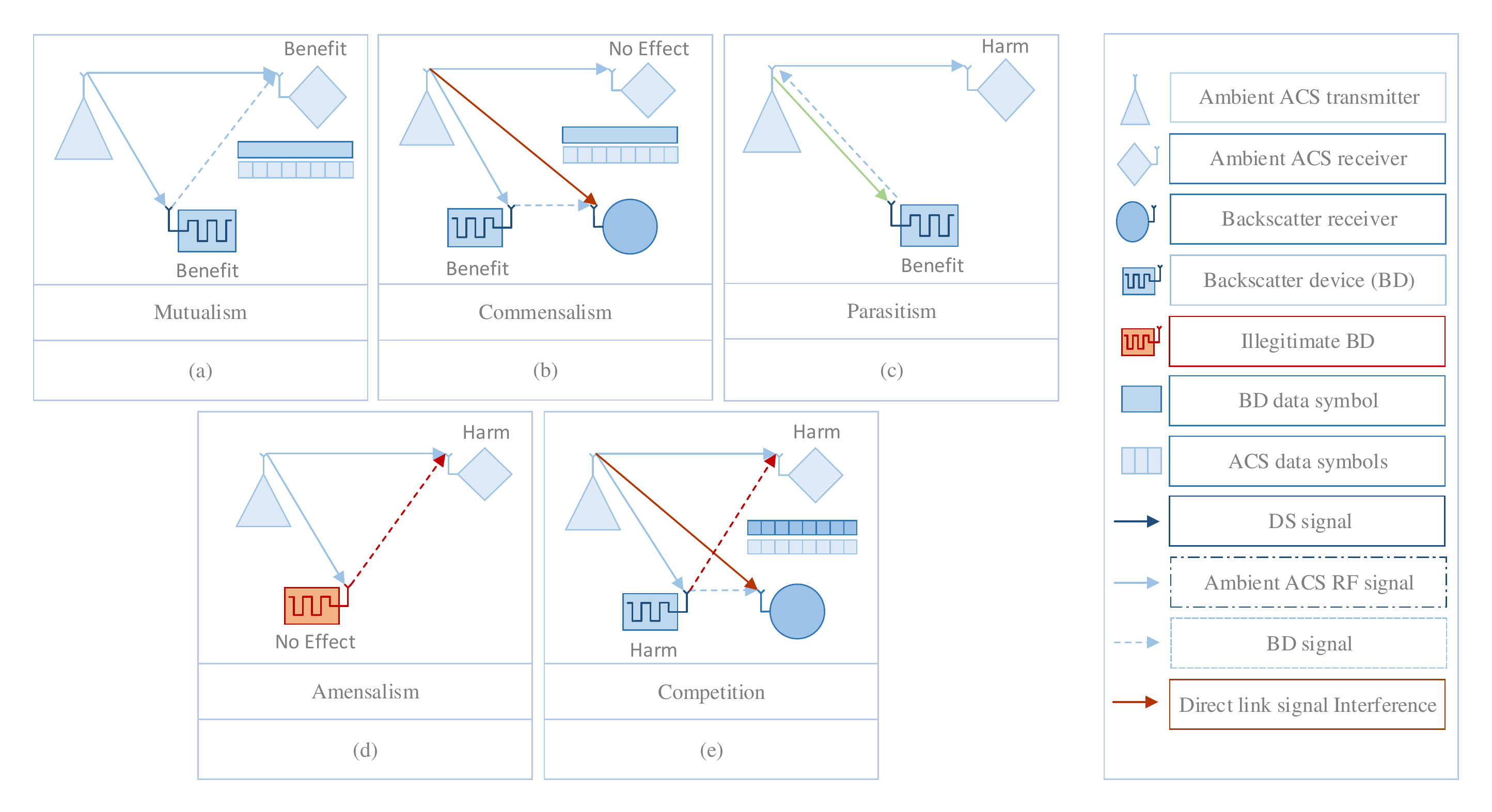}
    \caption{Symbiotic relationships between \ac{ACS} and \ac{PCS}.}
    \label{fig:Fig5}
\end{figure*}  

\section{Symbiotic Communication}
    This section discusses the architecture of the \ac{SRad} systems for \ac{SCm} with dependent coexistence between \ac{PCS} and \ac{ACS}. Different symbiotic relationships are explored while highlighting the recent research developments related to \ac{SCm}. Then, we present the techniques to overcome the performance of \ac{SCm}.

  \begin{table*}[ht]
  \centering
    \caption{Summary and Categorization of SCm Systems}
    \label{tab:TableII}
  \begin{center}
    \begin{tabular}{|m{0.06\linewidth} | m{0.25\linewidth}| m{0.07\linewidth}| m{0.075\linewidth}| m{0.06\linewidth}| m{0.075\linewidth}|}
    \hline \hline
   
    & & \multicolumn{4}{c}{\textbf{Symbiotic Relationships}} \vline \\ \cline{3-6}
     \multirow{-2}{*}{\textbf{Reference}} & \multirow{-2}{*}{\textbf{Design objective}} & \textbf{Mutualistic}&\textbf{Commensal}& \textbf{Parasitic}& \textbf{Competitive}\\ \hline \hline
     \cite{Long2020}& \ac{BD} transmission rate maximization and \ac{AT} power minimization.  & \textbf{--} & \cmark & \cmark & \textbf{--}\\ \hline
     \cite{Guo2019a}& \ac{BD} and \ac{AT} transmission rate maximization& \textbf{--} & \cmark & \cmark & \cmark \\ \hline
     \cite{Zhang2020DL} & Optimal \acp{BD} association &\textbf{--}&\textbf{--}&\cmark&\textbf{--} \\ \hline
     \cite{Liang2020}& \ac{BD} and \ac{AT} transmission rate maximization& \cmark& \textbf{--}&\textbf{--} & \textbf{--}\\ \hline
     \cite{Long2020a}& \ac{BD} transmission rate maximization and interference reduction at \ac{AR}  &\textbf{--}&\textbf{--} &\cmark&\textbf{--}\\ \hline
      \cite{Tuo2020}& Performance analysis of decode and forward relay network &\textbf{--}&\textbf{--}&\cmark&\textbf{--}\\ \hline
      \cite{Liu2020}&Interference free multiple \ac{BD} access. &\textbf{--}&\cmark&\textbf{--}&\textbf{--}\\ \hline
      \cite{Han2021}& Multiple \ac{BD} access& \textbf{--}&\textbf{--}&\cmark&\textbf{--}\\ \hline
      \cite{Guo2019}& \ac{BD} Resource allocation  &\textbf{--}&\cmark&\cmark&\textbf{--}\\ \hline
      \cite{Chu2020}& \ac{BD} Resource allocation &\textbf{--}&\cmark&\cmark&\textbf{--}\\ \hline
      \cite{Chen2020}& Stochastic transceiver design for multiple \ac{BD} &\textbf{--}&\textbf{--}&\cmark&\textbf{--}\\ \hline
      \cite{Zhang2019noma}&Outage probability and ergodic rate analysis of \ac{BD} &\textbf{--}&\cmark&\textbf{--}&\textbf{--}\\ \hline
      \cite{Li2019}& Secure beamforming and secrecy rate analysis of \ac{BD} &\textbf{--}&\cmark&\textbf{--}&\textbf{--}\\ \hline
      \cite{Long2019a}&Full-duplex \ac{BD} and \ac{AT} transmit power optimization &\textbf{--}&\textbf{--} &\cmark&\textbf{--}\\ \hline
      \cite{Long2019}& \ac{AT} and \ac{BD} transit power optimization &\textbf{--}&\textbf{--}&\cmark&\textbf{--}\\ \hline
      \cite{Liao2020}&\ac{BD} Resource allocation &\textbf{--}&\textbf{--}&\cmark&\textbf{--}\\ \hline
      \cite{Wu2019}& \ac{BD} transmission rate optimization &\textbf{--}&\textbf{--}&\cmark&\textbf{--}\\ \hline
      \cite{Dai2021}& Channel estimation and transmission rate analysis &\textbf{--}&\textbf{--}&\cmark&\textbf{--}\\ \hline
      \cite{Chen2021}&\ac{AT} transmit power minimization&\cmark&\textbf{--}&\textbf{--} &\textbf{--}\\ \hline
      \cite{Zhou2020}&\ac{AT} transmission power minimization &\textbf{--}&\cmark&\textbf{--}&\textbf{--}\\ \hline
      \cite{Zhang2020LISA}& \acp{BD} rate enhancement &\cmark&\textbf{--}&\textbf{--}&\textbf{--}\\ \hline
      \cite{Xu2021}& \ac{AT} transmit power minimization&\cmark &\textbf{--}&\textbf{--}&\textbf{--}\\ \hline
      \cite{Hua2020}& Performance enhancement of RIS assisted \ac{SRad} system by exploiting \ac{UAV} signals and trajectory &\textbf{--}&\cmark&\textbf{--}&\textbf{--}\\ \hline
    \end{tabular}
    \end{center}
\end{table*} 

            \subsection{Overview}
           The difference between conventional \ac{ABCmS} and \ac{SCm} is the different ways of radio resource sharing between the \acp{SRad} in coexistence through symbiotic relationships. These relationships are usually provide mutual benefits to both systems. Examples of symbiotic relationships are shown in Fig. \ref{fig:Fig5}.

            \subsection{Symbiotic Relationships}
            The \ac{SRad} system for \ac{SCm} with different symbiotic relationship scenarios is shown in Fig. \ref{fig:Fig5}(a), where \ac{ABCmS} performs the passive communication using the ambient signals transmitted by \ac{ACS} and both systems have mutualistic relationships. The system consists of three components: a transmitter and a receiver of \ac{ACS}, and \ac{BD}. The transmitter transmits the radio signal to transfer its information to the receiver, which is also re modulated and backscattered by the \ac{BD} for its data transmission. In this system, the receiver receives two signals: direct link high power signal from the transmitter and low power reflected signal from \ac{BD}. The receiver needs to detect the data of its own and \ac{BD}, while both systems are in mutualistic relationships, the \ac{BD} sends the data at a very low symbol rate compared to the transmitter's symbol rate, and it creates the multipath diversity at the receiver. Besides, the receiver first decodes its data and then applies successive interference cancellation to decode the data of \ac{BD}. The other symbiotic relationship scenarios for \ac{SCm} are also shown in Fig. \ref{fig:Fig5}. A commensal system is shown in Fig. \ref{fig:Fig5} (b), where the \ac{BD} gets the benefits by utilizing the ambient signals of \ac{ACS} for its communication and transmit at a low symbol rate to not affect the communication of \ac{ACS}. A full-duplex transmitter based parasitic \ac{SRad} system is shown in Fig. \ref{fig:Fig5} (c), in which the \ac{BD} transmits concurrently during the \ac{ACS} transmission to improve its communication but the performance of \ac{ACS} is degraded due to interference. An amensal \ac{SRad} scenario is shown in Fig. \ref{fig:Fig5} (d), where an illegitimate \ac{BD} creates the interference to harm the performance of the \ac{ACS} but it does not get any benefit or harm from the association. Lastly, a competition-based symbiotic relationship is shown in Fig. \ref{fig:Fig5} (e), where both system compete each other for improving their performance and \ac{BD} send at equal symbol rate as \ac{AT}.
            
            Several research studies have been done to exploit the aforementioned symbiotic relationships for the performance enhancement of \ac{SRad} systems. Authors in \cite{Guo2019a} investigate the resource allocation problem for maximizing the data rate in cooperative \ac{ABCm} based \ac{SRad} system in different fading states. Three symbiotic relationships including commensal, parasitic, and competitive, are proposed based on the transmission rate of \ac{ACS} and \ac{BD}. In a commensal relationship, \ac{ACS} tries to maximize its data rate without considering the performance of \ac{BD}. On the other hand in a parasitic relationship, \ac{BD} transmits with maximum reflection coefficient and at the same rate as \ac{ACS} to achieve the maximum data rate while sacrificing the transmission rate of \ac{ACS}. In the competition, both systems try to achieve the maximum transmission rate by competing for resources and harming each other. Lastly, the closed-form solution is derived for optimal power allocation for each symbiotic relationship under average power constraint. The extension of this work is proposed in \cite{Long2020}, where the authors analyze the maximum achievable rates of each system in both symbiotic relationships. Furthermore, the optimal beamforming is applied at the \ac{ACS} transmitter to optimize the rates of each system. 
         
            In \cite{Guo2019}, resource allocation problem is studied for \ac{SRad} system under fading channels. Commensal and parasitic symbiotic relations between \ac{ABCmS} and \ac{ACS} are investigated to overcome the spectrum growth problem, in which \ac{BD} transmits at different data rates compared to \ac{ACS} transmitter. Further, the maximum transmission rate of \ac{BD} and ergodic weighted sum rates of \ac{ACS} transmitter are achieved by jointly optimizing the reflection coefficient and transmit power of \ac{BD} and \ac{ACS} transmitter, respectively. The authors in \cite{Chu2020}, investigated the resource allocation problem in cooperative and non-cooperative \ac{SRad} systems and try to solve it with the finite blocklength channel codes under different transmission rates and symbol periods of \ac{BD}. Two optimization schemes are also proposed to minimize the transmission power of \ac{ACS} and maximize the energy efficiency of \ac{BD}.
               
            A decode and forward relay-based \ac{SRad} for \ac{ABCm} is investigated in \cite{Tuo2020}, where the relay is capable of simultaneous information and power transfer. The relay assists the \ac{ACS} and parasitic \ac{ABCmS} in forwarding their information to their destinations using power domain \ac{NOMA}. While observing the performance of both systems, it is concluded that this relaying mechanism achieves better throughput than the conventional at the cost of minor throughput degradation of \ac{ACS} system.
  
            In \cite{Liang2020}, a \ac{SRad} system with mutualism is proposed for \ac{ABCm} and \ac{CR}. Instead of doing active transmission, \ac{CR} backscatters the primary \ac{ACS} signals for secondary communication, and joint decoding is performed at the secondary receiver to decode the secondary and primary signals. Further investigations are made using a full-duplex secondary transmitter and \ac{RIS}, to improve the performance of primary \ac{ACS} and backscattered-\ac{CR} systems. Other than that in \cite{Long2020a}, \ac{CR} based \ac{ABCmS} is proposed as a \ac{SRad} system with an active load assisted \ac{BD} as a parasite in the \ac{ACS}. The authors considered this \ac{SRad} system as a secondary user and designed the beamforming vector to maximize the rate of \ac{CR} system along with the reflection gain of \ac{BD} while minimizing the interference to the primary user. 
            
            In a \ac{SRad} system, different radio systems can have multiple transmitters and receivers sharing the resources in a symbiotic manner. However, interference between the users and resource allocation are the critical issues that need to be addressed for the concurrent communication of multiple users with better performance. In \cite{Liu2020} authors investigated the interference issues in a multi \ac{BD} \ac{ABCm} based \ac{SRad} system, where multiple \acp{BD} transmit simultaneously to the \ac{ACS} receiver. The authors designed the coding algorithms with orthogonal code chips to enable interference-free communication for multiple \acp{BD}. Another multiple \ac{BD} \ac{SRad} system is proposed in \cite{Han2021} with distributive multiple \ac{BD} access scheme to overcome the frequent inter-system and inter-\ac{BD} coordination, where precoding is performed at each \ac{BD} by multiplying the data with random spreading code. Furthermore, an iterative-based algorithm is proposed to optimize the signal to interference plus noise ratio at \ac{BD} and minimize the transmission power of \ac{ACS} transmitter. A stochastic transceiver is designed in \cite{Chen2020} to solve the challenges faced by multiple \ac{BD} \ac{SRad} systems in the downlink transmission, that are inter-\ac{BD} interference and downlink interference. The authors also proposed a batch stochastic parallel decomposition algorithm as a solution to the stochastic multiple ratio fractional non-convex problems such as coverage analysis and acquiring the real-time tag's symbol information with minimum feedback signaling. Moreover, resource allocation problem is investigated in full-duplex \ac{SRad} network with \ac{NOMA} based \ac{ACS} system in \cite{Liao2020}. The \ac{ACS} transmitter has full-duplex functionalities and simultaneously sends the signals to \ac{ACS} receiver while receiving signals from \acp{BD}. Also, a \ac{NOMA} enhanced dynamic time division multiple access is proposed to enhance the spectral efficiency of the \ac{SRad} system.      
            In \cite{Zhang2020DL}, optimal user association problem is studied in multi-user \ac{SRad} system with network of \ac{ACS} users and \acp{BD} i.e., \ac{IoT} devices, to maximize the sum rate of all \acp{BD}. The main challenge in solving this problem is to estimate the channel of all \ac{ACS} users and \acp{BD} in real-time. The authors proposed two deep learning algorithms namely \ac{CDRL} learning and \ac{DDRL}  to infer the current \ac{CSI} utilizing the historical \ac{CSI} and concluded that both algorithms achieve optimal users association close to the one with perfect real-time \ac{CSI}. Furthermore, the \ac{DDRL} algorithm outperforms the \ac{CDRL} algorithm in terms of scalability for varying numbers of \acp{BD}, while the centralized algorithm needs less information than the distributed algorithms to solve the user association problem.  
            
            In \ac{SRad} system, backscattered signal is a very weak signal compared to direct link signal, which is similar to power domain \ac{NOMA}, where two users are which allows the multiple users to share the same spectrum resource with distinction in power, code or space domains \cite{Zhang2019noma}. Until now, the \ac{NOMA} systems are used to improve the spectral efficiency of \ac{ACS}; however, the spectrum efficiency can be enhanced further with the inclusion of the \ac{PCS} to support the passive \ac{IoT} systems. For examples, an \ac{ABCmS} can support its communication by developing a symbiotic relationship with \ac{NOMA} assisted \ac{ACS} e.g., cellular \ac{NOMA} system. Some studies have been conducted on \ac{SRad} systems with \ac{ABCmS} and \ac{NOMA} based \ac{ACS} \cite{Zhang2019noma, Li2019}. In \cite{Zhang2019noma}, a \ac{SRad} system consists of an \ac{ABCm} and \ac{NOMA} based \ac{ACS} system is presented, where \ac{BD} backscatters its data to the \ac{NOMA} users and the near user retrieve its data through successive interference cancellation. However, the far user treats the \ac{BD} signal as noise. The performance of the system is analyzed in terms of ergodic rates of \ac{BD} and \ac{NOMA} systems. A secure beamforming approach is proposed in \cite{Li2019} to enhance the secrecy of \ac{BD} in a multiple input-single-output \ac{NOMA} based \ac{SRad} system. The contributions of the aforementioned research studies related \ac{SCm} are summarized in Table \ref{tab:TableII} along with the classification in terms of symbiotic relationships that are used for the analysis.
            
\begin{figure*}[ht]
    \centering
    \includegraphics[width=\textwidth] {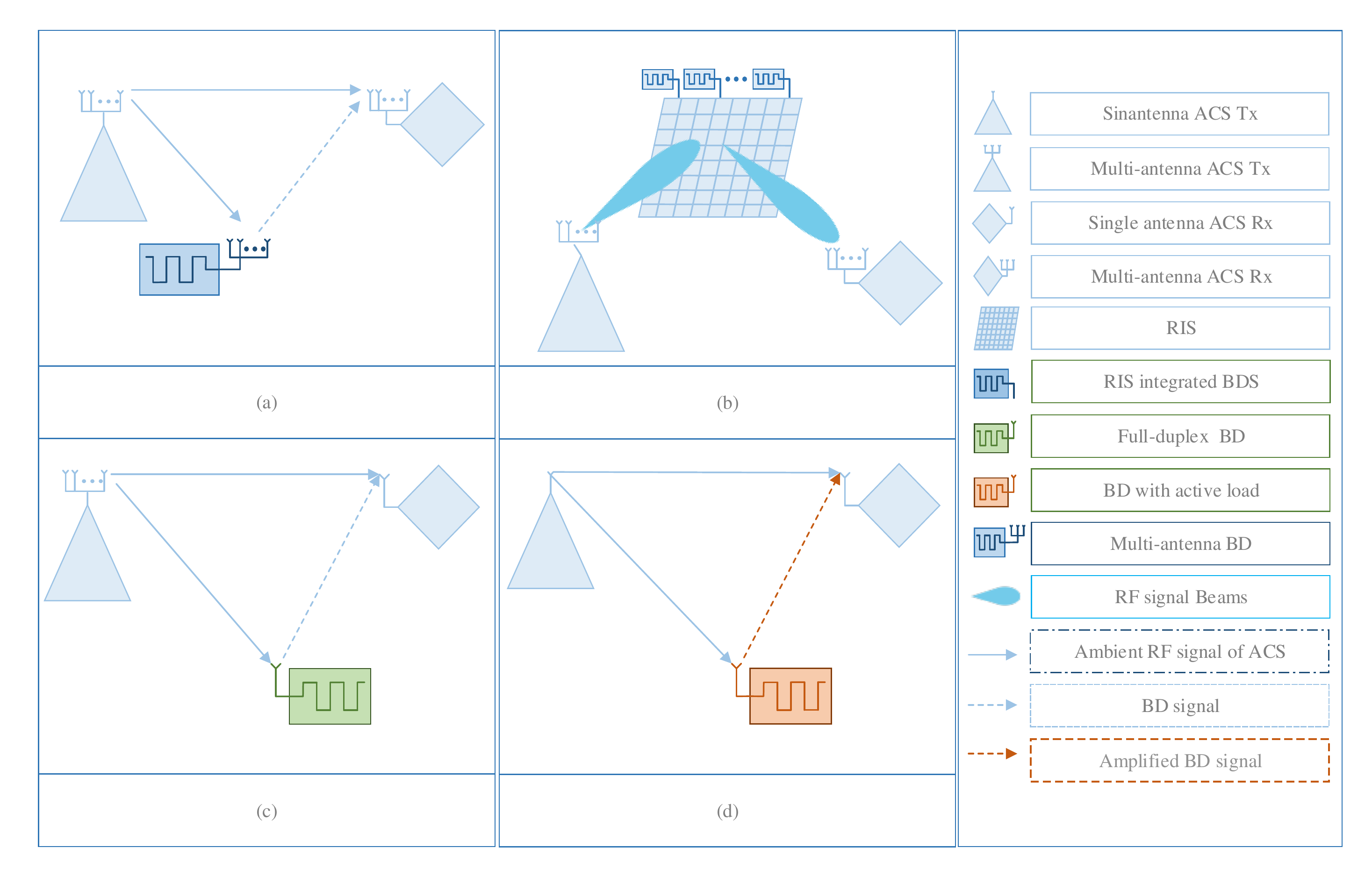}
    \caption{\ac{SRad} system performance enhancement using: (a) \ac{MIMO} \cite{Wu2019} (b) \ac{RIS} \cite{Zhang2020LISA} (c) full-duplex \ac{BD}\cite{Long2019a} (d) active-load assisted \ac{BD} \cite{Long2020a}. }
    \label{fig:Fig7}
\end{figure*}   
\subsection{Performance Enhancement}
            To improve the performance of the \ac{SRad} systems, different methods and technologies have been proposed such as \acl{MIMO}(\ac{MIMO}), full-duplex, \ac{RIS} and active-load assisted \ac{BD}. This presents the discussion on the implementation of these technologies in \acp{SRad} and their benefits.
               
            \subsubsection{\ac{MIMO} \ac{SRad} System} 
            Multi-antenna technology has been deployed widely to improve the performance of the \ac{ARS} through spatial diversity and spatial multiplexing. Therefore, to exploit the benefits of \ac{MIMO} technology for improving the performance of \ac{SRad} systems different studies have been conducted \cite{Wu2019, Dai2021}. An end-to-end \ac{MIMO} \ac{SRad} system is proposed in \cite{Wu2019}, where each component of the system is equipped with multiple antennas as shown in Fig. \ref{fig:Fig7}(a). An optimal beamforming design problem is investigated to maximize the transmission rate of \ac{BD} subject to the constraints of \ac{ACS} transmission rates. To solve this problem a locally optimal solution is proposed based on an exact penalty to obtain the capacity upper bound. A \ac{SRad} system based on symbiosis of \ac{ABCmS} and cell-free massive \ac{MIMO} systems is investigated in \cite{Dai2021}. Firstly, a two-step uplink training algorithm is also proposed to estimate the \ac{CSI} of the backscattered and direct links, wherein the first step is to estimate the direct link from the pilots received from \ac{ACS} transmitter without the involvement of \ac{BD} while in the second step both \ac{BD} and \ac{ACS} transmitters transmit for estimation of backscattered link. Secondly, low complex beamforming is exploited to derive the achievable transmission rates of both systems.
        
            \subsubsection{\Ac{RIS} Assisted \ac{SRad} System}
            \ac{RIS} or \ac{LIS} is a two-dimensional metasurface that consists of a large number of low-power scattering components. It is capable of reflecting the incident signals into desired directions through passive beamforming \cite{Dai2020}. Recently, due to its low power consumption and passive beamforming feature, it is considered as a promising solution to minimize the transmit power of \ac{ACS} while enhancing the achievable rate of \ac{BD} in a \ac{SRad} system  \cite{Chen2021,Zhou2020,Zhang2020LISA,Xu2021,Zhang2020a,Hua2020}.  A simple \ac{RIS} assisted \ac{SRad} is shown in Fig. \ref{fig:Fig7}(b), where multiple \acp{BD} are integrated to the \ac{RIS} and passive beamforming is approach is used to improve the communication link of \ac{ACS} link as well as to transfer the information of \ac{BD} to the receiver.
            
            In  \cite{Chen2021}, \ac{RIS} is deployed in a downlink \ac{SRad} system to enhance the performance of \ac{ACS} and multiple \acp{BD}, where the \acp{BD} face the severe blockage from the obstacles. Besides, the transmission power of the \ac{ACS} is minimized by jointly designing the active beamforming at \ac{ACS} transmitter and \ac{RIS} considering the signal-to-noise ratio and transmission rates of both systems as optimization constraints. A cooperative beamforming approach is proposed in \cite{Zhou2020} for a simple \ac{ABCm} based \ac{SRad} system. Wherein, \ac{RIS} transmits the data of \ac{IoT} device (i.e., connected to it through a wire) to \ac{ACS} receiver by modulating the incident signals while reconfiguring those signals to enhance the performance of \ac{ACS} receiver. \ac{RIS} can also transmit the data of multiple \acp{BD} connected to it via wired links to \ac{ABCm} receiver while assisting the wireless transmission of \ac{ACS}. To enhance the transmission rate of the \ac{ABCmS} without affecting the \ac{ACS} rate, a joint optimization approach for active and passive beamforming is studied in \cite{Zhang2020LISA}. Other works on \ac{RIS} assisted \ac{SRad} system with multi-user \ac{ACS} and a single \ac{IoT} device are investigated in \cite{Xu2021,Zhang2020a}. The \ac{IoT} device is connected to \ac{RIS} through wire, when \ac{RIS} reflects the \ac{ACS} transmission signals to its users, it simultaneously modulates the data of \ac{IoT} device and sends to \ac{IoT}. Furthermore, this study focuses on the \ac{ACS} transmission power minimization problem under the rate constraints of \ac{ACS} users and \ac{IoT} device, which is solved by jointly optimizing active and passive beamforming at base-station and \ac{RIS}, respectively. Authors in \cite{Hua2020}, investigated the \ac{RIS} assisted \ac{SRad} system under the coverage of \ac{UAV}. The \ac{RIS} modulate the signals transmitted by \ac{UAV} to transmit its data to the base station of terrestrial \ac{ACS} and in the meanwhile, it reconfigures the transmission link between \ac{UAV} and base station to improve the performance of \ac{UAV}. Further, the \ac{UAV} mobility is also exploited in this study to improve the transmission links between \ac{UAV}-base station and \ac{UAV}-\ac{RIS}. Particularly, to improve the error rate performance of \ac{RIS} through joint optimization of the \ac{UAV} trajectory and \ac{RIS} scheduling along with its phase shift matrix.        
            
            \subsubsection{Full-duplex \ac{SRad} System}
        
            Full-duplex technology is critical for spectral efficiency and reducing the latency of the system. However, it requires sophisticated and complex hardware for successive interference cancellation, which is not suitable for low power and low complex \ac{PCS} i.e., \ac{IoT}. In An \ac{ABCmS}, \ac{BD} also receives some control information to maintain proper communication, which creates latency in the system. A \ac{SRad} system with full-duplex \ac{BD} is shown in Fig. \ref{fig:Fig7}(c), where a \ac{BD} is capable of simultaneously transmitting and receiving data from the incident signal. To enable full-duplex \ac{ABCm} based \ac{SRad} system, a low complex full-duplex \ac{BD} design is proposed in \cite{Long2019a}. An extension of this work is proposed \cite{Long2019a}, where the authors optimized the transmit power of the \ac{ACS} along with the optimization of power splitting factor at \ac{BD}. The performance of the full-duplex \ac{SRad} system is measured in terms of transmission rates of \ac{BD}.

            \subsubsection{Active-load Assisted \ac{BD} \ac{SRad} System }
            One of the main reasons for the short range of \ac{PCS} is double-fading attenuation. The backscatter signal received at \ac{BR} is a low power signal compared to the interference signal \cite{Long2020a}. One way to solve this problem is to use \ac{RIS} at \ac{BD} and increase the signal power through passive beamforming \cite{Zhang2020LISA}. However, this is not possible in all cases, especially when the \ac{BD} is a small \ac{IoT} device or sensor. In this case, an active-load can be used at \ac{BD} to amplify and backscatter the incident signal through negative resistance \cite{bousquet20114}. Furthermore, it can reduce the dynamic range between the backscatter and direct link signal powers to increase the probability of correct signal detection at \ac{BD} with an additional cost of power to amplify the signal. This additional power consumption is still less than the power consumed by \ac{AT} with \ac{RF} chains \cite{khaledian2019active}. A \ac{SRad} system with active-load assisted \ac{BD} is shown in Fig. \ref{fig:Fig7}(d), where different symbiotic relationships can be possible between \ac{BD} and \ac{AT}. For instance, if the \ac{BD} transmit with equal symbol rate as \ac{AT} with power amplification to increase its transmission rate,  the performance of \ac{ACS} can be significantly affected. One the other, \ac{BD} can also assist the \ac{ACS} by providing a strong additional multipath, or it can relay the data of \ac{ACS} while backscattering its data in case of direct signal blockage at \ac{AR} \cite{Li2020}.   

 \begin{figure*}[ht]
    \centering
    \includegraphics[width=\textwidth] {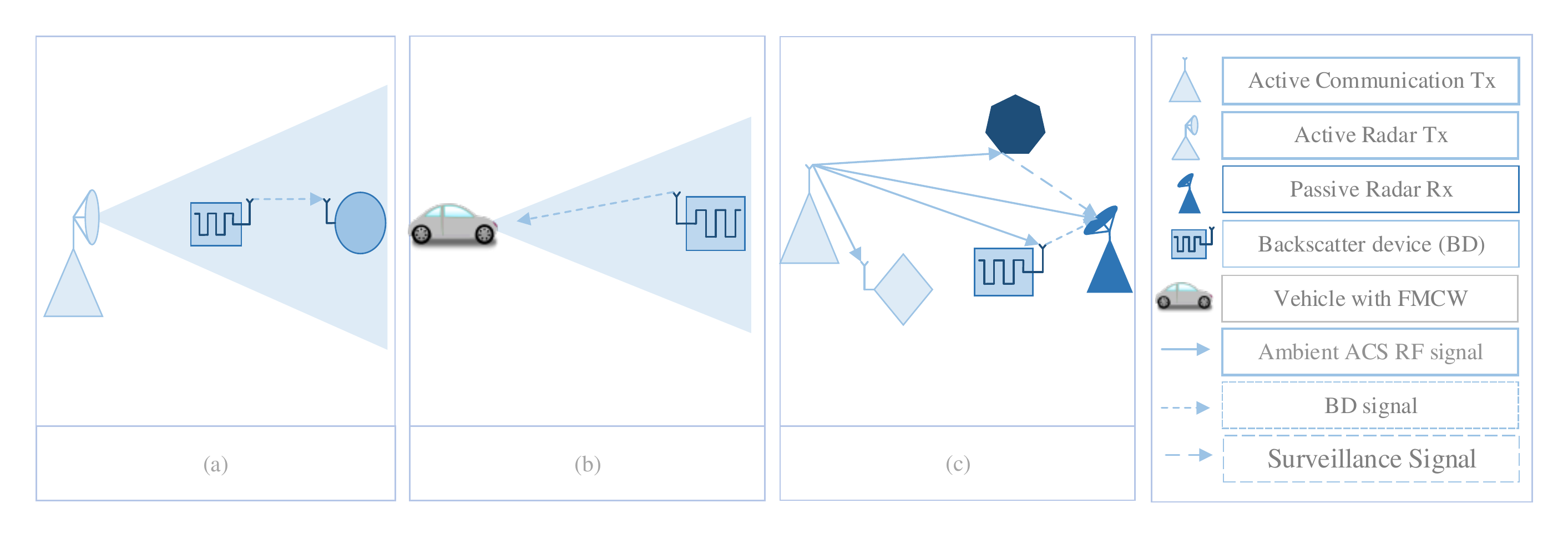}
    \caption{\ac{SSaC} Systems: (a)\ac{PCS} and monostatic \ac{ARDS}  (b) \ac{PCS} and \ac{ARDS} in vehicular scenario, (c) \ac{PCS} and \ac{PRDS}.}
    \label{fig:Fig6}
\end{figure*}     

\section{\ac{SRad} Enabling \ac{SRD}/\ac{SSaC}}
            In this section we first discuss the \ac{SRD}/\ac{SSaC} and symbiotic relationships. Afterwards, we discuss three use case scenarios of \ac{SRD} in which \ac{PRS} have dependent coexistence with \ac{ARDS} and \ac{PRDS}. 
            \subsection{Overview}
                A \ac{SRad} system can be a \ac{SRD}/\ac{SSaC} system if it performs both \ac{Radar} and communication functionalities. A common example of the \ac{SSaC} system is a \ac{PRS} that utilizes the ambient signals emitted active communication or radar systems (also known as illuminators of opportunity) for a target detection. The \acp{PRS} also termed as commensal \ac{Radar} systems because they do not affect the performance of the ambient signal source unlike \ac{PCS} as they do not transmit any information. Therefore, these systems are called  \cite{Mishra2014,Mishra2015}. As the \ac{Radar} and communication are different radio systems with their own identities and functionalities, they utilize the same radio resources and associated with each other on different terms. In literature, the radar and communication coexistence is presented with different names such as \textit{joint radar and communication}, \textit{integrated sensing and communication}, \textit{joint communication and radar}, \textit{joint sensing and communication}, and \textit{symbiotic sensing and communication} \cite{Wang2021sym}. Interested user are directed to the following surveys and tutorials \cite{Naik2018,Labib2016,Wang2021sym}.

        \subsection{Symbiotic Relationships}
            \acp{SRD}/\ac{SSaC} can also have different symbiotic relationships for sharing the radio resources which are expressed as follows:
         \begin{itemize}
            
            \item Mutualistic \ac{SRD}: A mutualistic relationship between a \ac{Radar} and a communication system occurs when both systems are associated in all aspects e.g., hardware design, spectrum, networks, etc., and promote each other for mutual benefits. They fully collaborate and coordinate with each other to achieve the highest performance gains.
         
            \item Commensal \ac{SRD}: Commensal symbiotic relationships can be found in \ac{SRD} systems with both \ac{ARDS} and \ac{PRS}. As mentioned earlier in \ac{PRS}, the \ac{Radar} system utilizes the ambient signals of \ac{ARS} for detection or sensing of the targets. In contrast to \ac{PRS}, \ac{ARDS} can develop a cooperative \ac{SRD} system with \ac{ACS}, where they exchange the information that helps in improving their performance without harming the other system.  
         
            \item Parasitic \ac{SRD}: The uncoordinated association between the \ac{ARDS} and \ac{ACS} leads to the parasitic symbiotic relationship, where one system shares the spectrum of the other without any coordination and degrades the performance of the other system by creating interference. For instance, \ac{ACS} utilizes the spectrum of \ac{ARDS} without coordination, when it transmits the communication signal, this signal interferes with the \ac{ARDS} and reduces its performance. 
         
            \item Neutral \ac{SRD}: In this type of \ac{SRD} systems, both \ac{ARDS} and \ac{ACS} use the same spectrum resources but in an opportunistic manner, which is possible if both systems share the information of their locations and the spectrum resources utilization. In this way, neither of the systems gets affected by the association but this \ac{SRD} is not spectral efficient.
         
            \item Competitive \ac{SRD}: When \ac{ARDS} and \ac{ACS} share the same radio resource under a competitive symbiotic relationship, each one tries to enhance its performance without taking into account the other system. For instance, when \ac{ACS} shares the spectrum with \ac{ARDS}, the \ac{Radar} system transmits with high power to enhance its performance and degrades the communication of \ac{ACS}. 
        \end{itemize}
        
        \ac{PCS} can also utilize the ambient \ac{RF} signals generated by the \ac{Radar} systems to perform their communication without affecting its functionality. We discuss three use case scenarios for symbiosis between \ac{PCS} and \ac{ARDS} to enable both \ac{Radar} and passive communication. In the first case, a \ac{SRad} system based on the commensal relationship between \ac{PCS} and \ac{ARDS} is considered as shown in Fig. \ref{fig:Fig6} (a), where \ac{BD} utilizes the signal transmitted by the \ac{Radar} transmitter for communication. In this \ac{SRad} system, the major limitation is the short and intermittent pulses of the \ac{Radar} systems that require strict synchronization at \ac{BD}. In \cite{Cnaan-On2014}, a frequency modulated continuous wave \ac{Radar} waveform is used to support both localization and \ac{BD} communication. A framework is also designed for joint processing of \ac{BD} communication and \ac{Radar} signals and a line coding-based filter is used for clutter removal to detect the target of interest within clutter return. This work is extended for distributed sensing and communication in \cite{Cnaan-On2015}. In the second use case, vehicle-to-vehicle/vehicle-to-infrastructure scenarios are considered, illustrated in Fig. \ref{fig:Fig6}(b), where a car transmits the frequency modulated continuous wave for \ac{SSaC} functions. There is also a \ac{BD} connect to an \ac{IoT} device in the environment that wants to send its information to the car. When the signal strikes the \ac{BD} antenna, it remodulates the signal with its information. A clutter return is received at the receiver of the vehicle, from which the sensing and communication information is extracted using the clutter filter \cite{Cnaan-On2015}. Last but not least, a \ac{SRad} system with a \ac{BD} and a \ac{PRDS} is presented in  Fig. \ref{fig:Fig6}(c), where the \ac{PRDS} use the signal of the \ac{ACS} for performing the sensing/\ac{Radar} functions and receiving the information of the \ac{BD}. \ac{PRDS} receiver get both the direct link and the \ac{BD} signals, initially, it measures the direct signal and then correlates it with the backscattered signal to obtain the required information for both the communication and \ac{Radar}. These use cases are discussed simplify the explanation of \ac{SRD}/\ac{SSaC} systems. Readers interested in \ac{PRDS} are directed to study the tutorial given in \cite{Kuschel2019passive}.

\section{Applications}
        The interdependent coexistence through multiple symbiotic relationships make the \ac{SRad} systems a strong candidate to support the massive connectivity of \ac{IoT} devices in \ac{6G} \cite{Zhang20196GVision, Liang2020}. In this section, we discuss the applications of \ac{SRad} systems and divide them into two categories: (i) human-centric; (ii) machine-centric. The human-centric applications are the ones that impact the living standards of humans, e.g., E-health, whereas the machine-centric are related to the communication between sensors and devices helping in a process to enhance the efficiency, e.g., manufacturing.

    \subsection{Human Centric} These applications are designed to support the healthy life and well-being of humans. Although \ac{SRad} can support all the applications related to \ac{IoT}, yet there are a few applications that have achieved significant developments and play a critical role in human life.
        \paragraph{E-Health}
          Health monitoring through different devices and sensors is one of the critical applications of \ac{IoT} in \ac{6G} \cite{Janjua2020}. New implantable sensors and devices are used to observe the temperature, oxygen level, blood pressure, etc. Due to the low power and limitation of computational capability, these devices cannot use conventional wireless communication technologies because their power consumption is very high. On the other hand, \ac{SRad} consumes less power compared to the aforementioned technologies and enables the energy harvesting from ambient signals of \ac{ACS} systems. Furthermore, information transfer between these medical devices and \ac{ARS} is very easy as they share the same radio resources and infrastructure.
        \paragraph{Wearable Devices}
            The second human-centric application of \ac{IoT} is the wearable devices, which collect information from the human body and transfer it over the internet. These devices are also low-power devices and require low-power wireless connectivity for information transfer. Besides, providing a continuous connection to these devices under mobility is another critical challenge. \ac{SRad} can provide these devices low power communication and continuous connectivity through a symbiosis between the \ac{PCS} and long rage \ac{ARS}. 
        \paragraph{Ambient Assisted Living}
            The elderly population is increasing rapidly, and most of them prefer to stay at home. Due to old age, people suffer from loneliness,  and diseases such as dementia, which creates difficulty for them to locate their daily use things, e.g., medicine kits. To assist the elderly person in their daily activities, a \ac{SRad} based \ac{IoT} network can be developed in the home to support positioning, localization, and environment information. The household devices equipped with \ac{BD} can send their information to \ac{WiFi} access point during its communication to the elderly person's cell phone in the uplink/downlink. The \ac{SRad} can work in different symbiotic relationships with the \ac{WiFi} or cell phone to improve each other performance.  
    \subsection{Machine Centric} Proliferation of \ac{IoT} devices and sensors in the industrial sector, agriculture, energy, and several other sectors is increasing continuously. These devices need to communicate and transfer their information among each other or to an information center. This is an open research problem to meet the radio resource needs through the existing wireless systems. \ac{SRad} systems can be an enabler for such application areas to support massive connectivity within the limited spectrum through symbiosis with ambient \ac{ACS}.     
        \paragraph{Manufacturing}
            In the manufacturing process, \ac{IoT} devices, i.e., sensors, are deployed for machine-to-machine and machine-to-human communication, enhancing the information flow, centralized control, and intelligent monitoring. Accommodating these devices with wireless connectivity can be achieved by developing a \ac{SRad} network. Through this, the large number of devices can use the signals of the ambient \ac{ACS} through symbiotic relationships as well as play a part in improving its performance through multipath diversity.
        \paragraph{Agriculture}
            Collecting the information about the environment and utilizing it to support agriculture can be realized through \ac{IoT} based ecosystem, where the devices are placed far distant apart and require low power and long-range communication. \ac{SRad} systems in commensal relationship to the \ac{TV} and FM base station can provide these services to the devices with low power consumption compared to existing wireless technologies, e.g., LoRa. 
        \paragraph{Vehicle-to-Everything}
            Autonomous driving is one of the critical application areas of \ac{6G}, in which the vehicle needs to communicate to the other vehicles as well as to infrastructure for safe driving and prevent accidents. Several sensors are installed in the vehicle to obtain the critical information from the environment through \ac{Radar}/sensing. If cellular V2X or dedicated short-range communication (DSRC)  are supporting vehicular communication,  signals coming from these systems can be utilized to enable passive communication between vehicles and infrastructure through \ac{SRad}. Autonomous vehicles can share their mobility information with other vehicles, or the road safety sign boards can backscatter the critical information to the vehicles. Thus, \ac{SRad} provides the spectrum efficiency with massive connectivity for vehicle-to-everything communication.     
     
\section{Open problems and Future Directions}
        Although different research studies on \ac{SRad} systems provide different insights and techniques for the development of these systems, yet there are many open research problems to solve before \ac{SRad} become mature. Here are some of the critical challenges highlighted to direct the further research on \ac{SRad} systems for \ac{SCm}. 
            \subsection{Transmitter Design}
                In a \ac{SRad} system, the transmitter design of one system can affect the other system depending on the symbiotic relationship. Therefore, we first look at the challenges related to transmitter design and also discuss the possible solutions.    
                \paragraph{Joint Waveform Design} 
                    Unlike the \ac{ACS}, where the carrier is a continuous wave signal, \ac{PCS} uses the data modulated \ac{RF} signal of \ac{ACS} and remodulate it to send its information. In different symbiotic relationships, both systems can have comparatively different data rates. For instance, in a competition to achieve better throughput both systems can transmit at the same symbol rate, which can affect the performance of both systems negatively. Therefore, joint waveform design can be a potential solution to support \ac{SCm} with better performance and high spectral efficiency. 
                \paragraph{Joint Modulation Design}
                     If the \ac{PCS} uses the same modulation scheme as \ac{ACS} with the same symbol rate, it makes the detection complicated at the receiver. To prevent this situation a joint modulation design can be considered that can allow using higher-order modulations coherently while improving the performance of both systems.   
                \paragraph{Smart Multi-antenna Design}
                     Although multi-antenna technology and their implementation protocols as well algorithms have been explored extensively for \ac{ACS} performance enhancement; however, these are not directly applicable to \ac{PCS} due to their high power consumption and complexity. Therefore, novel smart antenna designs for \ac{PCS} with low complex algorithms for communication and data aggregation are required to achieve performance improvements.
                \paragraph{Hardware design of \ac{PCS} transmitter}
                    Synchronization of \ac{ACS} and \ac{PCS} transmitters' symbols is a critical problem. The circuit used in \acp{ACS} for timing and phase recovery is power-consuming. There is a need for low power circuit and low complexity synchronization algorithms for synchronization at RF level in \ac{PCS} transmitter. Active load and \ac{MIMO} are the available options, but their power efficiency is less.
                    The full-duplex capability can allow the \ac{PCS} transmitter to harvest energy and enhance the power of the signal using the harvested energy. It can provide other benefits as well such as simultaneous signal detection and reflection.

    \subsection{Receiver Design}
                \paragraph{Signal Detection} 
                    
                    The direct path signal is several times stronger than the backscattered signal, which is not visible within the dynamic range of the ADC controller; thus, results in a higher packet error rate. Different methods can be used to reduce the error rate: i) Before decoding the \ac{PCS} signal first eliminate the direct path signal; ii) Use beamforming and directing the null of the antenna towards the strong signal can allow the reception of the \ac{PCS} signal. Then, traditional ADC can sample the weak \ac{PCS} signal and a non-coherent receiver can be used to detect it. Further performance improvement can be achieved using coding e.g., Hadamard codes, which allows the signal detection in very noisy and unreliable channels.
                    
                    \paragraph{Signal Processing under Practical Limitations}
                    Robust signal processing techniques are required in real-time implementation of \ac{PCS} because the \ac{PCS} signal is very weak, and the availability of \ac{CSI} is not certain at all times. In addition, \ac{CSI} becomes old and needs to be reattained after some time. Besides, \ac{NOMA}-assisted SRad systems require the successive interference cancellations for a large number of \ac{PCS} devices, which become computationally complex.  
                    
                    \paragraph{Sporadic Transmission}
                    In practical scenarios, \acp{PCS} use the ambient signal sporadically for transmission of data and ambient signal tracking at \ac{PCS} devices is critical for its successful transmission. One solution is the synchronization between the \ac{PCS} device and \ac{ACS} transmitter to have the pre-knowledge of ambient signal transmission time. This requires the design and implementation of low power and low complex algorithms at \ac{PCS} device.
                    \paragraph{Downlink signaling/feedback and uplink channel}
                    The low power and simple processing constraints limit the capabilities of the \ac{PCS} device as a receiver. However, the downlink control channel is necessary to design a physical layer protocol for \ac{PCS}. Besides, the uplink channel design is also important for feedback and data transfer. A non-coherent receiver design is comparatively simple than the coherent one but with low performance. On the other hand, an energy detector requires a much larger symbol length of \ac{PCS} signal than \ac{ACS} signal to average out the underlying modulated signal. 
    \subsection{Network Design and Management}
                \paragraph{\ac{ACS} Source Selection and User Association}
                    More \ac{ACS} transmitters with multiple channels can improve the energy harvesting efficiency and provide more spectrum holes for concurrent active data transmissions. This complicates the whole system in terms of trade-offs among energy harvesting, backscattering, active transmission, and channel selection.
                    Furthermore, in the presence of a lot of ambient signals coming from different \ac{ACS}, but the selection of an appropriate source is critical for the development of sustainable \ac{PCS}. The propagation and availability of RF signals vary in environments. For instance, the indoor environment has different channel characteristics than outdoor environments. Selecting the suitable \ac{ACS} resource can have a strong impact on the performance of the \ac{PCS} and the overall performance of \ac{SCm}. Moreover, user association considering the performance of both \ac{PCS} and \ac{ACS} users as well as increasing the performance of the overall system. In the existing studies, it is assumed that the single \ac{PCS} users is associated with the single \ac{ACS} user and utilize the resources of that specific user for \ac{SCm}, which is not practical because most of the time number of \ac{PCS} user is less than or greater to the \ac{ACS} users. In these scenarios, more than one \ac{PCS} users can be associated to the single \ac{ACS} user, or one \ac{PCS} user can utilize the signal of multiple \ac{ACS} users for \ac{SCm}.    
                
    \subsection{Channel Modelling and Estimation}
            In \ac{SRad} systems for communication, modeling the channels of both \ac{ACS} and \ac{PCS} are essential to compensate for the combined effect of channels on both systems. Also, \ac{PCS} has a different propagation channel than the \ac{ACS} in the sense that its channel is cascaded and doubly faded. Moreover, channel estimation is also critical especially for \ac{PCS} to detect the weak signal at the receiver in the presence of interference from the direct link signal of \ac{ACS}. The conventional schemes for \ac{ACS} are not directly applicable to the \ac{SRad} systems due to their high complexity. To design a proper \ac{SCm}, it is highly desired to estimate the channel accurately. Furthermore, \ac{RIS} is being used in \ac{SRad} to improve its performance; however, channel estimation in \ac{RIS} is already an open problem, with the addition of \acp{PCS} devices to the \ac{RIS} increase the severeness of the problem.                 
    \subsection{Security and Privacy}
            In SRad, both systems have different symbiotic relations in terms of resource utilization, disrupting the communication of one source can affect the other system directly. Besides an illegitimate \ac{ACS} can also try to acquire the sensitive information of the \ac{PCS} device and play an amensal role and harm the performance of the \ac{PCS} users. Securing the data of \ac{PCS} is vital due to its critical and sensitive data applications, particularly in healthcare. Conventional cryptography-based encryption and authentication schemes designed for \ac{ACS} devices need high computationally complex algorithm implementation. Thus, the design of a low complexity scheme is necessary for such a device. Additionally, physical layer security techniques can be utilized to provide security and privacy in \ac{PCS}. Authentication of ambient signal prior utilization is needed to secure the \ac{PCS} signals. Furthermore, the broadcast nature of backscattered signals makes them vulnerable to security attacks. 
    \subsection{Performance Analysis and Metrics}
            In \ac{SRad} systems the radio system shares the resources in multiple dimensions e.g., spectrum, energy, infrastructure, new performance metrics need to be designed to measure the performance in multiple dimensions, unlike the conventional performance metrics. In the commensal system, \ac{PCS} device has a very high data rate which can be considered as additional multipath to the ambient system. The ambient system can utilize this to improve its performance. The overall capacity of the system increases with the addition of the \acp{PCS} devices to the \ac{ACS} in a mutualism symbiotic relationship. This is an interesting research area to investigate that how much throughput or capacity gains can be achieved with the addition of the \ac{PCS} device to the \ac{ACS}. 
            
\section{Conclusion}
    To meet the needs for future wireless systems within limited radio resources is challenging, especially with the proliferation of wireless applications. The continuous development of \ac{IoT} devices and sensors is making the problem worse and existing radio resource sharing mechanisms are failing to fulfill the need of radio systems. Thus, new paradigms should be explored for spectrum sharing and coexistence. In this survey, we highlight the importance of using
    \ac{PRS} i.e., \ac{PCS} and \ac{PRDS} along with the \ac{ARS}, and discuss a bio-inspired mechanism called symbiosis for resource sharing and coexistence of in radio ecosystem to enable \ac{SCm} and \ac{SRD}. There are two types of symbiosis, i.e., obligatory and facultative depending on the radio system dependency and mapped it to coexistence. If one of more radio systems in symbiosis depend on each other completely, e.g., \ac{ARS} and \ac{PRS} then symbiosis is obligatory (i.e., dependent coexistence). On the other hand, if the radio systems can survive separately and have symbiosis as well then the resultant symbiosis is facultative (i.e., independent coexistence). Our focus is on dependent coexistence due to \ac{PRS}. Then, different symbiotic relationships between dissimilar \ac{SRad} systems are analyzed. It is observed that these relationships may have significantly varying realizations ranging from the cooperation of the symbionts to competition between them for the radio resources. The performance enhancement techniques for \ac{SRad} systems are presented along with the different use case scenarios. Applications of \ac{SRad} communication from human and machine-centric visions are also described for future wireless networks. We foresee \ac{SRad} as a potential candidate for efficient spectrum sharing between \ac{PRS} and \ac{ARS} as well as for the coordination between the dissimilar \ac{ARS} in the future.




\begin{IEEEbiographynophoto}{Muhammad Bilal Janjua}
(Student
Member, IEEE) received the B.E. and M.Sc.
degrees in electrical engineering from the University of Lahore, Lahore,
Pakistan, in 2012 and 2017, respectively. He is
currently pursuing the Ph.D. degree with the Com-
munications, Signal Processing, and Networking
Center (CoSiNC), Istanbul Medipol University,
Turkey. He is also a member of the Communica-
tions, Signal Processing, and Networking Center
(CoSiNC), Istanbul Medipol University. His research interest includes Internet of Things, symbiotic radio, and backscatter communication for future wireless networks.  
\end{IEEEbiographynophoto}
\vfill
\begin{IEEEbiographynophoto}{H\"{u}seyin Arslan}
(Fellow, IEEE) received the
B.S. degree from Middle East Technical Univer-
sity (METU), Ankara, Turkey, in 1992, and the
M.S. and Ph.D. degrees from Southern Methodist
University (SMU), Dallas, TX, USA, in 1994 and
1998, respectively.

From January 1998 to August 2002, he was
with the Research Group of Ericsson, where he
was involved with several projects related to 2G
and 3G wireless communication systems. Since
August 2002, he has been with the Department of Electrical Engineering,
University of South Florida, where he is currently a Professor. In Decem-
ber 2013, he joined Istanbul Medipol University, to found the Engineering
College, where he has worked as the Dean of the School of Engineering
and Natural Sciences. In addition, he has worked as a part-time Consultant
for various companies and institutions, including Anritsu Company and
the Scientific and Technological Research Council of Turkey. He has been
collaborating extensively with key national and international industrial part-
ners and his research has generated significant interest in companies, such
as InterDigital, Anritsu, NTT DoCoMo, Raytheon, Honeywell, Keysight
Technologies. Collaborations and feedback from industry partners has sig-
nificantly influenced his research. In addition to his research activities, he has also contributed to wireless communication education. He has integrated the
outcomes of his research into education which led him to develop a number of
courses at the University of South Florida. He has developed a unique Wire-
less Systems Laboratory course (funded by the National Science Foundation
and Keysight Technologies), where he was able to teach not only the theory
but also the practical aspects of wireless communication system with the
most contemporary test and measurement equipment. He conducts research
in wireless systems, with emphasis on the physical and medium access layers
of communications. His current research interests include 5G and beyond
radio access technologies, physical layer security, interference management
(avoidance, awareness, and cancellation), cognitive radio, multi-carrier wire-
less technologies (beyond OFDM), dynamic spectrum access, coexistence
issues, non-terrestrial communications (high altitude platforms), joint radar
(sensing), and communication designs.

Dr. Arslan has served as a Technical Program Committee Member,
the General Chair, the Technical Program Committee Chair, a Session and
Symposium Organizer, and the Workshop Chair for several IEEE confer-
ences. He has also served as a member for the Editorial Board for the IEEE
TRANSACTIONS ON COMMUNICATIONS, the IEEE TRANSACTIONS ON COGNITIVE
COMMUNICATIONS AND NETWORKING (TCCN), and several other scholarly jour-
nals by Elsevier, Hindawi, and Wiley Publishing. He is a member of the
Editorial Board for the IEEE COMMUNICATIONS SURVEYS AND TUTORIALS and
the Sensors Journal.
\end{IEEEbiographynophoto}
\end{document}